\documentclass[11pt, a4paper]{article}
\usepackage[absolute]{textpos}
\usepackage[top=3cm,bottom=3cm,left=2cm,right=2cm]{geometry}
\usepackage{amsmath,amssymb,mathtools,dsfont,emptypage,graphicx,dcolumn,bm,booktabs,colortbl,collcell,enumerate,float,verbatim,multirow,xparse,slashed}
\usepackage{cite}
\usepackage{tabularx}
\usepackage{multirow}
\usepackage{graphicx}
\usepackage{subcaption}
\usepackage{appendix}
\usepackage{booktabs} 
\usepackage{colortbl}
\usepackage{collcell}
\usepackage{xparse}
\usepackage{relsize}

\usepackage[labelfont=bf, labelsep=period]{caption}

\usepackage[compat=1.1.0]{tikz-feynman}
\usepackage{contour}

\def\hc{\text{h.c.}}

\def\z2{$\mathbb{Z}_2$}
\def\321{$\mathrm{SU(3)_c} \times \mathrm{SU(2)_L} \times \mathrm{U(1)_Y}$}

\def\Tnuc			{T_{\text{nuc}}}
\def\Treh			{T_{\text{reh}}}
\def\Tfo			{T_{\text{fo}}}

\def\UV			{\mathsmaller{\text{UV}}}
\def\GW			{{\mathsmaller{\text{GW}}}}
\def\DM			{{\mathsmaller{\text{DM}}}}
\def\D			{{\mathsmaller{\text{D}}}}
\def\EW			{{\mathsmaller{\text{EW}}}}
\def\LHC			{{\mathsmaller{\text{LHC}}}}
\def\SN			{{\mathsmaller{\text{SN}}}}
\def\NS			{{\mathsmaller{\text{NS}}}}
\def\Pl			{\mathsmaller{\text{Pl}}}

\def\min{\text{min}}

\definecolor{myred}{cmyk}{0,1,1,0.55}
\definecolor{mygreen}{rgb}{0.27, 0.64, 0.48}
\definecolor{myblue}{cmyk}{0.8, 0.4, 0, 0.2}
\definecolor{mygray}{gray}{.95}
\definecolor{niceorange}{rgb}{0.9, 0.3, 0.2}
\definecolor{nicepurple}{rgb}{0.7, 0.0, 0.4}
\usepackage[linktocpage,colorlinks=true,urlcolor=mygreen,linkcolor=red,citecolor=blue]{hyperref}

\begin{document}

\begin{center}
{\bf\LARGE Nanohertz gravitational waves\\ [3 mm]
from the baryon-dark matter coincidence}\\
[5mm]
\renewcommand*{\thefootnote}{\fnsymbol{footnote}}
Alessia Musumeci $^{a}$ \footnote{\href{mailto:alessia.musumeci@tum.de}{alessia.musumeci@tum.de}},
Jacopo Nava$^{b,c,d}$
\footnote{\href{mailto:jacopo.nava@vub.be}{jacopo.nava@vub.be}},
Silvia Pascoli $^{b,c}$ \footnote{\href{mailto:silvia.pascoli@unibo.it}{silvia.pascoli@unibo.it}},
and
Filippo Sala$^{b,c}$
\footnote{\href{mailto:f.sala@unibo.it}{f.sala@unibo.it}, FS is on leave of absence from LPTHE, CNRS and Sorbonne Universit\'e, Paris, France.} 
\\
$^{a}$\,{\it \small Physik-Department, Technische Universit{\"{a}}t M{\"{u}}nchen, James-Franck-Stra{\ss}e 1, 85748 Garching, Germany}\\
$^{b}$\,{\it \small Dipartimento di Fisica e Astronomia, Università di Bologna, via Irnerio 46, 40126 Bologna, Italy} \\
$^{c}$\,{\it \small INFN, Sezione di Bologna, viale Berti Pichat 6/2, 40127, Bologna, Italy}\\
$^d$\,{\it \small Theoretische Natuurkunde and IIHE, Vrije Universiteit Brussel,\\
\& The International Solvay Institutes, Pleinlaan 2, B-1050 Brussels, Belgium }
\end{center}

\begin{center}
    {\bf Abstract}
\end{center}
    \noindent 
    The nanohertz gravitational waves (GW) observed by pulsar timing arrays may originate from a cosmological first-order phase transition (PT) at $\sim$ 100 MeV. Taking this possibility seriously motivates the question: why 100 MeV?
    We point out that a PT at exactly those scales is predicted by the generation of the baryon asymmetry from a dark asymmetry via resonant neutron-dark matter oscillations, and we show that this PT can induce an observable GW signal compatibly with all experimental constraints. This proposal predicts dark matter self-interactions close to their observational upper limits and lowers the maximal expected mass of neutron stars. Independently of GW, this baryogenesis mechanism is tested by searches for missing-energy at the LHC and for neutron decays. We keep the model consistent with big-bang nucleosynthesis by adding heavy neutral leptons below 100 MeV, which generate neutrino masses and can induce further experimental tests.
    \\
\hrule

\tableofcontents

\renewcommand*{\thefootnote}{\arabic{footnote}}
\setcounter{footnote}{0}

\section{Introduction}

In 2023 pulsar timing array experiments (PTA)~\cite{NANOGrav:2023gor, EPTA:2023fyk,Reardon:2023gzh,Xu:2023wog} reported the first observation of a gravitational wave (GW) signal compatible with being a stochastic GW background (SGWB), and hence the first one whose origin could possibly be cosmological~\cite{NANOGrav:2023hvm}. 
The other possibility for its origin is astrophysical, from mergers of super-massive black hole (SMBH) binaries ~\cite{NANOGrav:2023hfp}. While many analyses find that a cosmological origin fits data slightly better than an astrophysical one (e.g.~\cite{NANOGrav:2023hvm,Figueroa:2023zhu,Winkler:2024olr}), such assessments so far depend on modeling assumptions and astrophysical priors.

Viable cosmological sources for the observed nHz GW include cosmic strings, domain walls and strong first order phase transitions (PT) and they all require physics beyond the standard model (BSM).
This adds to the importance to tell apart a cosmological from an astrophysical origin of the signal. A robust way to do so is to rely on the same nHz GW, for example via the precise spectral shape of the signals and via their anisotropies (see e.g.~\cite{Depta:2024ykq}), although more progress is needed to make the most of upcoming data in this respect~\cite{Konstandin:2024fyo,Domcke:2025esw,Konstandin:2025ifn}. Another approach is to look for other signals of the putative cosmological source. This is the approach we take in this paper, focusing on PT.

\medskip

If the nHz GW signal was due to a PT, then that PT should reheat the universe at hundreds of MeV~\cite{NANOGrav:2023hvm} and thus be associated with BSM at those scales. Having determined the PT properties needed to explain the signal, the following step is to determine their implications for the sub-GeV particle physics models inducing the PT, see e.g. the recent~\cite{Costa:2025csj,Balan:2025uke,Bringmann:2026xcx}. In our opinion, a compelling next step down this path is to answer the question: why 100 MeV?

An intriguing answer would be to link the PT scale with QCD. The SM predicts that the QCD PT is a crossover~\cite{Aoki:2006we}, hence not generating GW.
Although BSM could make the QCD PT first order~\cite{Schwarz:2009ii,Chatrchyan:2025wop}, whether or not it could make it as strong and slow as required by the observed nHz GW depends on the model~\cite{Gao:2024fhm,Chatrchyan:2025wop}, motivating more work in this direction. 
We are not aware of other (successful) attempts to motivate the $O(100)$~MeV scale of the PT, that are independent of the nHz GW signal itself.

\medskip

In this paper, we explore the origin of a PT at $O(100)$~MeV from the baryon - dark matter (DM) coincidence problem. The measured baryon and DM energy densities are within a factor of 5 of each other, $\rho_\DM \simeq 5 \rho_b$, today and back to the earliest times observed. On one hand, $\rho_b$ is exponentially sensitive to $O(1)$ UV parameters, because dimensional transmutation (i.e. the strong coupling running from the UV until it induces confinement in the IR) sets the proton mass, on the other hand $\rho_\DM$ is sensitive to parameters that depend on its unknown production mechanism and are a priori unrelated to those setting $\rho_b$. The measured relation $\rho_\DM \simeq 5 \rho_b$ then is an extremely puzzling coincidence, because it requires a double tuning in the baryon and DM parameter space.

This puzzle can be addressed if DM is asymmetric like baryons, and a mechanism either generates the two asymmetries at the same time (`cogenesis') or transfers a previously generated DM asymmetry into a baryon one (`darkgenesis'), see~\cite{Petraki:2013wwa,Zurek:2013wia} for reviews.\footnote{These mechanisms actually relate the DM and baryon number densities. To reproduce $\rho_\DM \simeq 5 \rho_b$ they also need a relation between DM and proton masses, and they either simply posit it or achieve it with additional ingredients.
We are aware of only one mechanism addressing the coincidence problem  directly in terms of energy densities, the relaxation mechanism of~\cite{Brzeminski:2023wza,Banerjee:2024xhn}, but we are not aware of ways to connect it to nHz GW.}
While some cogenesis and darkgenesis models require a first order PT, they do not need it to take place at a scale that explains the nHz GW signal. A darkgenesis model that instead requires a BSM PT precisely at $O(100)$~MeV, first proposed in 2018~\cite{Bringmann:2018sbs}, partly transfers a dark asymmetry to the baryon one via resonant oscillations of the neutron in a dark sector partner lighter by one part in $10^4$-$10^5$, to be identified with DM. 
Ref.~\cite{Bringmann:2018sbs} also explored the connection of this mechanism with the neutron lifetime anomaly, which we do not pursue here. It did not explore its connection with nHz GW, which we carry out in this paper. Requiring that the PT has the properties needed to explain the PTA signal adds constraints on the model of~\cite{Bringmann:2018sbs}, and could a-priori close the region where it successfully addresses the baryon-DM coincidence.
We anticipate that a region of parameter space non-trivially survives, where the baryon-DM coincidence is addressed and observable nHz GWs are predicted, and where other observables could further test the model, see Fig.~\ref{fig:parameterspace}.

\medskip

This paper is organised as follows.  In Sec.~\ref{sec:baryonasymmetry} we introduce the effective interaction responsible for neutron-DM mixing and describe how the resulting oscillations can transfer a primordial dark asymmetry to the visible sector, thereby generating the observed baryon asymmetry of the universe (BAU). In Sec.~\ref{sec:potential} we build the scalar potential responsible for the dark PT, including finite-temperature corrections and the one-loop running of the couplings. The resulting PT dynamics and the associated GW signal are computed in Sec.~\ref{sec:Gravsignal}.
In Sec.~\ref{sec:constraints} we study the main cosmological, astrophysical, and collider constraints that affect the scenario and determine the viable parameter space. In Sec.~\ref{sec:BBN} we discuss the thermal history of the dark sector and present an explicit realization consistent with the cosmological bounds. Finally, we summarize our results and discuss future prospects in Sec.~\ref{sec:summary}.

\section{Generation of the baryon asymmetry}
\label{sec:baryonasymmetry}

In this section we describe the interactions responsible for the origin of the DM-neutron mixing, and the resulting transfer of the DM asymmetry to the visible sector.

\subsection{Dark matter - neutron oscillations}

We consider DM to be a Dirac fermion $\chi$ carrying unit baryon number, with mass $m_\chi$, charged under a dark $U(1)_\D$ gauge symmetry mediated by a dark photon $V$.
The interaction connecting $\chi$ to the SM up and down quarks, $u$ and $d$, arises from the higher dimensional operator~\cite{Cline:2018ami}
\begin{equation}
\label{eq:effectiveop7}
\mathcal{O}_{\rm eff} =
\frac{1}{\Lambda_7^3} \, \Phi \,
\left(\overline{u^c} P_R d \right)
\left(\overline{\chi} P_R d \right),
\end{equation}
where $\Phi$ is the complex scalar field responsible for $U(1)_\D$ spontaneous breaking and $\Lambda_7$ denotes the scale of heavy fields that have been integrated out. 
We will comment about UV completions of Eq.~(\ref{eq:effectiveop7}) in Sec.~\ref{sec:accelerator}.
When the scalar field $\Phi$ acquires a non-zero vacuum expectation value (VEV)$\langle \phi \rangle$, this operator generates an effective interaction between the dark fermion and the SM quarks.

Below the QCD confinement scale and the $\Phi$ PT, Eq. (\ref{eq:effectiveop7}) leads to an off-diagonal mass term between the neutron and the DM, given by 
\begin{equation}
\label{eq:neutronDMmixing}
\mathcal{L}_{\text{mix}} = -\delta m\, \bar{n}\chi + \text{h.c.},
\end{equation}
with
\begin{equation}\label{eq:deltam}
\delta m =\beta  \frac{\langle \phi \rangle}{\Lambda_7^3}\,,
 \end{equation}
where $ \beta=0.014 \, \text{GeV}^3$ from the lattice matrix element $\left\langle 0 \left| udd \right| n \right\rangle$ \cite{Aoki:2017puj}.
This mixing induces oscillations between $\chi$ and the neutrons, allowing a pre-existing dark asymmetry stored in the $\chi$ sector to be partially transferred to the visible baryons.
The dark asymmetry has to be generated at temperatures above the PT inducing the mixing in Eq.~(\ref{eq:neutronDMmixing}), but other than that we remain agnostic about its origin.
We will sketch an explicit example for the origin of the dark asymmetry in Sec.~\ref{sec:washout}.

We now discuss how the mixing term in Eq.~\eqref{eq:neutronDMmixing}
induces $\chi-n$ oscillations in the early Universe and allows the transfer of the dark asymmetry to the visible sector, following~\cite{Bringmann:2018sbs}. The dynamics of the $(n,\chi)$ system can be described by the $2\times2$ Hamiltonian
\begin{equation}\label{eq:mixingmatrix}
H = \begin{pmatrix}
m_n +\Delta E_n (T)  & \delta m \\
\delta m & m_\chi+ \Delta E_\chi (T) \,
\end{pmatrix}\,,
\end{equation}
where $\Delta E_n$ and $\Delta E_\chi$ are the thermal mass shifts of the neutron and of the DM respectively. We extract the former from~\cite{Eletsky:1997dw} and we estimate $\Delta E_\chi=g ^2 T_\gamma'^2/(8 m_\chi)$ from the dark photon contribution to the $\chi$ self energy, where $g $ is the gauge coupling of the dark $U(1)_\D$ and $T_\gamma'$ the temperature of the dark sector. We anticipate that we will later complement the model in such a way that $T_\gamma' = T_\gamma$, where $T_\gamma$ is the temperature of the SM photons.
The mixing angle $\theta$, obtained by diagonalizing $H$, reads
\begin{equation}
\tan\,(2\theta) = \frac{2\delta m}{\Delta m + \Delta E_n - \Delta E_\chi} \equiv \frac{2\delta m}{\delta E}\,,
\end{equation}
where $\Delta m \equiv m_n - m_\chi$, $\delta E = \Delta m + \Delta E_n - \Delta E_\chi$, and the difference between the eigenvalues $\omega_{1,2}$ of $H$ is given by $
|\delta\omega| = \sqrt{(\delta E)^2 + 4 \delta m^2}$.
To find the efficiency of $\chi - n$ oscillations, we start with the initial state \( \psi(0) = |\chi \rangle \) and evolve it with the Hamiltonian Eq. (\ref{eq:mixingmatrix}), obtaining

\begin{equation}
|\psi(t)\rangle = e^{-i(\omega_1 + \omega_2)t/2} \Bigg( \Big(\cos \frac{\delta \omega}{2}t - i \cos 2\theta \sin\frac{\delta \omega}{2}t \Big)|\chi \rangle - i \sin 2\theta \sin \frac{\delta \omega}{2} t |n \rangle \Bigg)\,,
\end{equation}
therefore the probability of $\chi$ to oscillate into a neutron is
\begin{equation}
    P_n (t)=\sin^2 (2\theta)\sin^2 \left(\frac{\delta\omega\, t}{2}\right)\,.
    \label{eq:Pnt}
\end{equation}

\subsection{Dark-to-baryon asymmetry transfer in the early universe}
\label{sec:darkbaryon}
Because of the large rate of interactions $\Gamma_n$ of neutrons
on heat bath pions, however, there may not be time for
a full oscillation.
One should then carry out a time average, over the short time scale $1/\Gamma_n$,
\begin{equation}\label{eq:averageprob}
    \bar{P}_n=\Gamma_n \int_0^{\infty}dt\, e^{-\Gamma_n t}P_n(t)=\frac{2 \delta m^2}{\delta \omega^2 +\Gamma_n^2}\,.
\end{equation}
The exponential damping accounts for the scatterings of neutrons with the thermal bath, which destroy the phase coherence between the $\chi$ and $n$ states. Hence, the averaged probability $\bar P_n$ can be interpreted as the probability for a $\chi$ particle to convert into a neutron during one scattering time.

The neutron scattering rate $\Gamma_n$ on the pions in the thermal bath is given by $\Gamma_n= n_\pi \langle \sigma_{n \pi} v_\pi\rangle$, with $n_\pi$ the pion thermal number density, including a factor of 3 due to isospin, $\langle v_\pi\rangle =\sqrt{8T \,/ (\pi \, m_\pi)}$ and $\sigma_{n\pi}\simeq 0.1/m_\pi^2 \simeq 2$mb~\cite{Fettes:1998ud,RuizdeElvira:2017stg}, with $m_\pi \simeq 140$ MeV the pion mass.
The rate of production of neutrons via oscillations per $\chi$ particle is given by $\Gamma_{\text{osc}} = \overline{P_n} \Gamma_n$.
Similarly, the production of \( \chi \) per neutron must proceed with the same rate $\Gamma_{\text{osc}} $, therefore the number density of $\chi$ can be written as
\begin{equation} \label{eq:evolution}
\dot{n}_\chi + 3H n_\chi= -\Gamma_{\text{osc}} (n_\chi - n_n)\,.
\end{equation}
Since in our scenario the total baryon number is conserved, the equation controlling the fraction 
$f = n_n/(n_n + n_\chi)$ of neutrons that are converted by DM is  given by
\begin{equation}
\dot{f}=\Gamma_{\text{osc}}(1-2f)\,, 
\end{equation}
which is solved by
\begin{equation} \label{eq:Asymmetry}
    f=\frac{1}{2}\left(1-\text{Exp}\Big(-2 \int dt \,\Gamma_{\text{osc}} (t)\Big)\right) \\
    \simeq\frac{1}{2}\left(1-\text{Exp}\Big(-2 \int \frac{dT}{T} \frac{\Gamma_n(T) \bar{P}_n(T)}{H(T)}\Big)\right)\,.
\end{equation}
Since $m_n \simeq m_\chi$ then $\Omega_{b}/\Omega_{\chi} = f/(1-f)$, so that
the observed baryon over DM abundance $\Omega_{b}/\Omega_{\chi} \simeq 0.19$~\cite{Planck:2018vyg} implies $f \simeq 0.16$.

As shown in Eq.~\eqref{eq:Asymmetry}, the final baryon abundance is controlled by the time integral of the oscillation rate. The integrand encodes the interplay between the oscillation probability and the Hubble rate $H^2(T)=\rho_\text{rad} (T)/(3 M_{\Pl}^2)$ in radiation domination, where $M_{\Pl} \simeq 2.4 \times 10^{18}$ GeV is the reduced Planck scale.
The quantity $\Gamma_n \bar{P}_n/H$ exhibits a sharp peak at the resonance temperature $T_\text{res}$, defined by the condition $\delta E(T_\text{res}) = 0$, where $|\delta\omega| = 2\delta m$ and the $\chi - n$ mixing is maximal (we verify that our calculation of $\Gamma_n \bar{P}_n/H$ reproduces Fig.~2 of~\cite{Bringmann:2018sbs}).
The generation of the baryon asymmetry is completely dominated by the resonant region.

\begin{figure}[t]
        \centering
        \includegraphics[width = 0.7\textwidth]{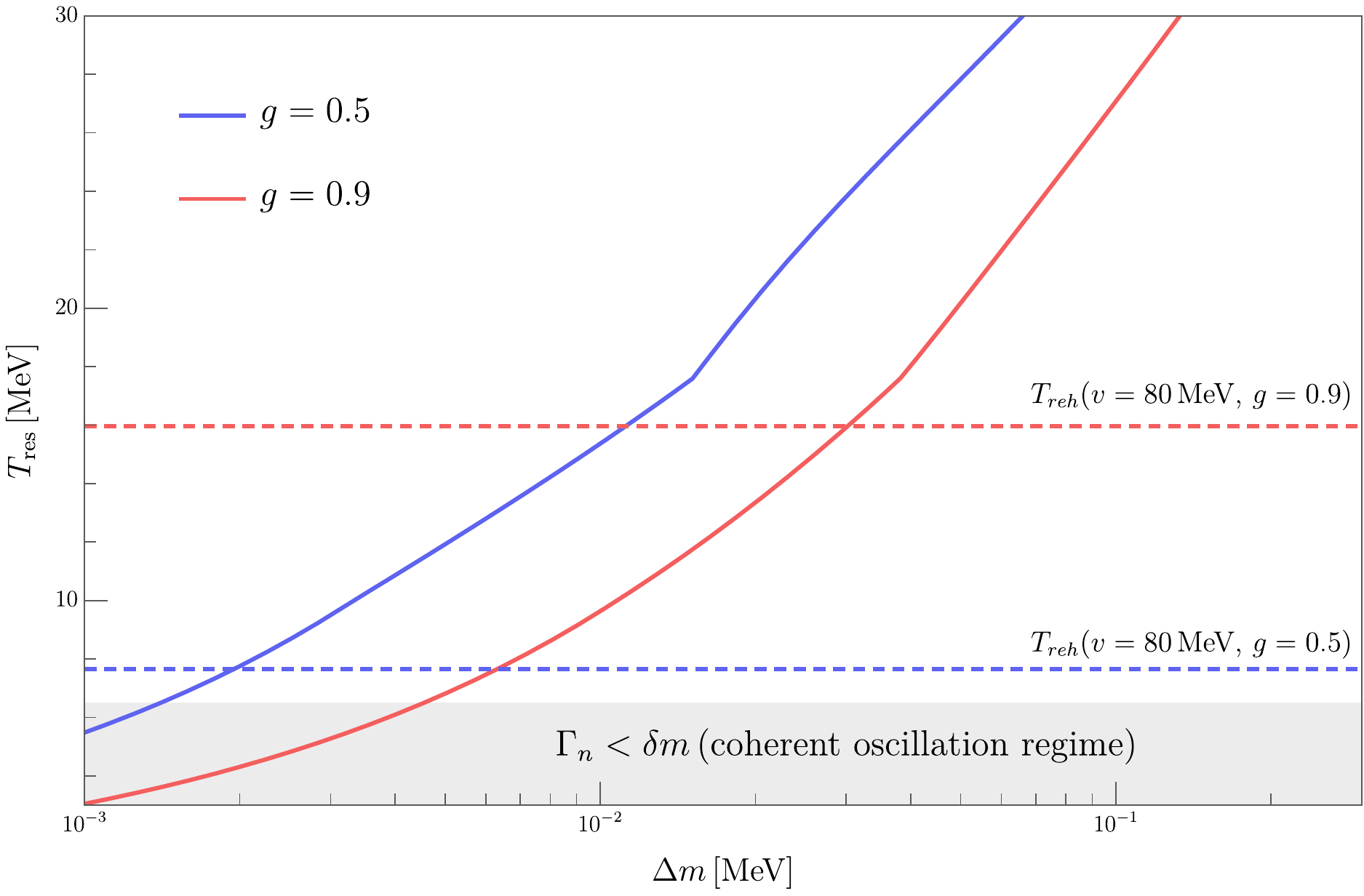}
        \caption{\textbf{Solid lines:} Resonance temperature for the neutron-DM oscillation defined as $\delta E (T_{\text{res}})=0$, for the benchmark values of the gauge coupling $g = 0.5$ (blue) and $g=0.9$ (red) and scalar field VEV $v=80$ MeV. \textbf{Dashed lines:} Reheating temperature $T_{\rm reh}$ for the same values of $g$ and $v$. When $T_\text{reh} < T_\text{res}$ the dark-to-baryon asymmetry transfer is suppressed. \textbf{Gray shaded region:} $\Gamma_n < \delta m$; in this region the fast-decoherence approximation that we adopted breaks down and coherent oscillations between dark matter and neutron become relevant, the computation should be then carried out using the density matrix formalism.
    }
                \label{fig:TResonance}

\end{figure}

We show in Fig. \ref{fig:TResonance} the value of the resonance temperature for benchmark values of $g=0.5,0.9$ and VEV $v=80$ MeV. 
In order for the baryogenesis mechanism to operate efficiently, after the $\Phi$ PT the Universe must reheat at a temperature $T_{\rm reh}>T_\text{res}$, ensuring that the plasma goes through the resonant regime in which $\chi$-$n$ oscillations are efficient.
To illustrate this requirement, in Fig.~\ref{fig:TResonance} we also display the reheating temperature $T_{\rm reh}$, calculated as described in Sec.~\ref{sec:potential}.
The condition $T_{\rm reh}>T_\text{res}$ then translates into an upper limit on $\Delta m$. Indeed, $T_\text{res}$ grows with increasing $\Delta m$, because $\delta E(T_\text{res})=0$ implies that larger thermal corrections $\Delta E_{n,\chi}(T_\text{res})$ are needed to cancel a larger $\Delta m$.

We finally comment on the validity of our analytical calculation of the dark-to-baryon asymmetry transfer. A fully rigorous treatment of oscillations in the presence of scattering would have required the use of the density matrix formalism~\cite{Cirelli:2011ac}, which tracks the off-diagonal correlations induced by oscillations and their damping due to interactions with the thermal bath. 
However, in the regime of interactions larger than oscillations $\Gamma_n > \delta \omega/2$ (see Eq.(\ref{eq:Pnt})) oscillations are efficiently decohered, the off-diagonal components of the density matrix are strongly suppressed, and the quantum kinetic equations reduce to the classical rate equation in Eq.~(\ref{eq:evolution}) which we have used, as explicitly verified in~\cite{Bringmann:2018sbs}. 
The fast interaction condition $\Gamma_n > \delta \omega/2$ translates, on resonance, to $\Gamma_n > \delta m$, which implies
\begin{equation}
T \gtrsim 6.5\,\mathrm{MeV}
\end{equation}
independently of $v$ for the parameters of our interest (because $\Gamma_n \propto n_\pi$ which is exponential in $T$).
We shade in gray the region where this condition is not satisfied and one would need to go beyond our analytical decohered treatment of Eq.~(\ref{eq:evolution}), by implementing a full density-matrix solution that  takes into account coherent oscillations.

\subsection{Parameter space of successful baryogenesis}
We now explore the relevant parameter space for the generation of the baryon asymmetry in the $\delta m - \Delta m$ plane.
The qualitative behaviour can be understood as follows. 
The oscillation probability is suppressed by the damping rate $\Gamma_n$, as described in Eq. (\ref{eq:averageprob}), 
and in the fast-interaction regime the averaged probability for DM-to-neutron conversion reads $\bar{P}_n \simeq 2 \delta m^2/ \Gamma_n$.
 Increasing the mass splitting $\Delta m$ shifts the resonance temperature $T_{\text{res}}$ to higher values, as shown in Fig.~\ref{fig:TResonance}. At higher temperatures, the pion density is higher, which increases the neutron scattering rate $\Gamma_n$. This increased damping suppresses oscillation-driven baryon production via $\bar{P}_n \simeq 2 \delta m^2/ \Gamma_n$. Consequently, in order to maintain a fixed baryon asymmetry, larger values of the mixing parameter $\delta m$ are required as $\Delta m$ increases.

The parameter space is subject to a number of experimental constraints. The strongest bound on the mass splitting $\Delta m$ arises from the stability of ${}^9\text{Be}$ nuclei, which requires $\Delta m < 1.665~\text{MeV}$~\cite{Fornal:2018eol}.
In addition, the UV cutoff scale is constrained as $\Lambda_7 < 1.9$~TeV by our reinterpretation, carried out in Sec.~\ref{sec:accelerator}, of monojet plus missing transverse energy searches (MET) at the LHC~\cite{ATLAS:2024vqf,Hiller:2026osz}. Using Eq.~(\ref{eq:deltam}), this limit can be translated into an upper bound on the mixing parameter $\delta m$ at fixed $v$. Finally, the $\delta m-\Delta m$ plane is constrained by null searches for the decay $n \to \chi + \gamma$~\cite{Tang:2018eln}.

\begin{figure}[t]
        \centering
        \includegraphics[width = 0.8\textwidth]{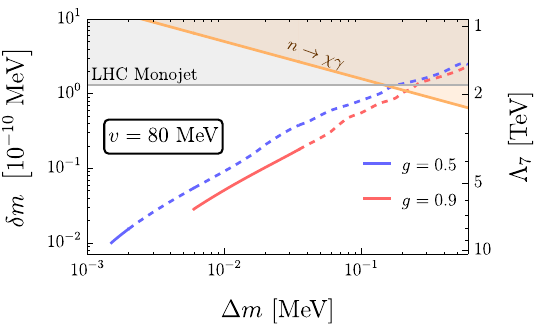}
        \caption{Solid lines yield the observed baryon asymmetry in the $\delta m- \Delta m$ plane for benchmark values of the dark gauge coupling $g=0.5,0.9$ from Eq. (\ref{eq:Asymmetry}). The dashed extensions correspond to the regime where the resonance temperature satisfies $T_{\rm res} > T_{\rm reh}$. In this regime the standard treatment of the oscillation dynamics remains valid, but the cosmological history is not consistent with the PT scenario assumed in this work. The solid lines are truncated when $\Delta m$ is so small to lower $T_\text{res}$ to the point $\Gamma_n(T_\text{res}) < \delta m$, where neutron-DM oscillations become coherent and the approach used to derive Eq.~(\ref{eq:Asymmetry}) breaks down.  The orange area is excluded from $n \to \chi \gamma$ null searches~\cite{Tang:2018eln}, the gray area is excluded by monojet+MET searches at the LHC \cite{ATLAS:2024vqf,Hiller:2026osz}, interpreted in Sec.~\ref{sec:accelerator} as a bound on the scale $\Lambda_7$ in Eq.~(\ref{eq:effectiveop7}), which we translate to a limit on $\delta m$ for the benchmark value of the $\Phi$ VEV $v=$ 80 MeV.}
       \label{fig:Baryogenesis_ParameterSpace}
\end{figure}

The resulting parameter space is shown in Fig.~\ref{fig:Baryogenesis_ParameterSpace}. The colored curves correspond to parameter values that reproduce the observed baryon asymmetry for the benchmark choices of $g = 0.5$ and $g=0.9$, which as we will see encompass the region of our interest generating observable nHz GW.
We stop the curves at $\Delta m$ so small that the temperature of dark-to-baryon asymmetry transfer, $T_\text{res}$, is small enough to induce $\Gamma_n < \delta m$, where our decohered analytical treatment is not valid. We turn the curves from solid to dashed at $\Delta  m$ so large that $T_\text{res} > T_\text{reh}$.
The orange region is excluded by searches for $n \to \chi \gamma$, while the gray region is ruled out by monojet+MET searches for a benchmark scalar VEV $v = 80\,\text{MeV}$.

From Fig.~\ref{fig:Baryogenesis_ParameterSpace} we observe that the observed BAU can be generated for $\Delta m \lesssim 0.2~\text{MeV}$, well below the bound from nuclear stability. However, once we will further require that baryogenesis occurs consistently within the PT framework discussed in Sec. \ref{sec:potential} and with the generation of nHz GW, the viable parameter space will be significantly restricted. This will establish a non-trivial link between the baryogenesis mechanism and the dynamics of the PT.

\section{Effective Potential}
\label{sec:potential}

\subsection{Scale-invariant Potential}
\label{sec:SIpotential}
In contrast to the setup of Ref.~\cite{Bringmann:2018sbs}, where the dark $U(1)_\D$ symmetry is broken by a scalar potential with an explicit mass scale, we assume classical scale invariance in the dark sector and generate the scalar VEV radiatively via the Coleman-Weinberg (CW) mechanism~\cite{Coleman:1973jx}. 
Our choice is motivated by the fact that classically scale-invariant potentials naturally lead to strongly supercooled FOPTs \cite{Randall:2006py, Konstandin:2011dr}, 
enhancing the resulting GW signal and thereby matching the features favored by the PTA signal interpretation in terms of a cosmological FOPT \cite{NANOGrav:2023hvm}. Such setups have been extensively explored in the literature as viable explanations of the observed nHz GW signal~\cite{Goncalves:2025uwh,Balan:2025uke,Bringmann:2026xcx}.

Concretely, we consider the scale-invariant Lagrangian
\begin{equation}
\mathcal{L}_{\rm dark} = -\frac{1}{4}F'_{\mu\nu}F'^{\mu\nu}-\frac{\epsilon}{2}F'_{\mu\nu}F^{\mu\nu}
+ |D_\mu \Phi|^2 - \lambda |\Phi|^4 \,,
\label{eq:Ldark}
\end{equation}
where $D_\mu=\partial_\mu - i g V_\mu$, with $V_\mu$ the dark photon associated with the $U(1)_\D$ gauge symmetry. We also allow a kinetic mixing $\epsilon$ between the dark photon and the SM photon. 
At tree level the scalar potential is therefore given by
\begin{equation}
V_{\rm tree}(\Phi)=\lambda |\Phi|^4\,,
\label{eq:scaleinv}
\end{equation}
and the symmetry breaking scale is generated radiatively.

We can now compute the one-loop corrections to the potential à la CW. We decompose $\Phi$ in terms of its real and imaginary components, expanding around the background field $\phi_c$:

\begin{equation}
    \Phi=\frac{\phi+i\varphi_i+\phi_c}{\sqrt{2}}\,.
\end{equation}
The one-loop effective potential at $T=0$, in the $\overline{\mathrm{MS}}$ renormalization scheme and Landau gauge, reads:
\begin{equation}\label{eq:oneloop}\begin{split}
    V_{\mathrm{eff}}^{T=0}(\phi_c)=\frac{\lambda}{4} \phi_c^4 + \frac{1}{64\pi^2}\biggl[3m_{V}(\phi_c)^4\biggl(\log{\frac{m_{V}^2(\phi_c)}{\mu^2}}-\frac{5}{6} \biggr)\\+m_\phi^4(\phi_c)\biggl(\log{\frac{m_\phi^2(\phi_c)}{\mu^2}}-\frac{3}{2} \biggr)+m_{\phi_i}^4(\phi_c)\biggl(\log{\frac{m_{\phi_i}^2(\phi_c)}{\mu^2}}-\frac{3}{2} \biggr) \biggr]\,,
    \end{split}
\end{equation}
where $\mu$ is the renormalization scale and the field-dependent masses for the real and imaginary components of $\Phi$ and the dark photon are
\begin{align}
    m_{\phi}^2(\phi_c)=3\lambda \phi_{c}^{2}\,,\\
    m_{\phi_i}^2(\phi_c)=\lambda \phi_{c}^{2} \,,\\
    m_{V}^{2}(\phi_c)=g^2\phi_c^2\,.
\end{align}

\begin{figure}[t!]
  \centering
    \includegraphics[width=0.48 \textwidth]{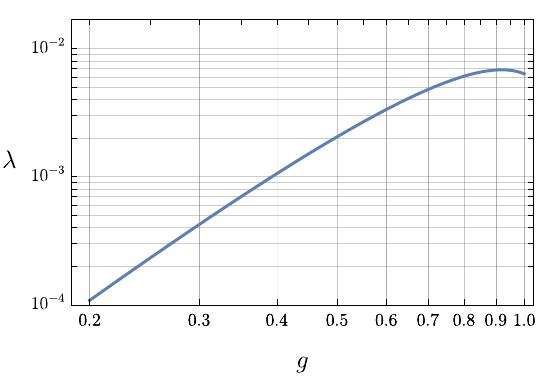}
    \includegraphics[width=0.48 \textwidth]{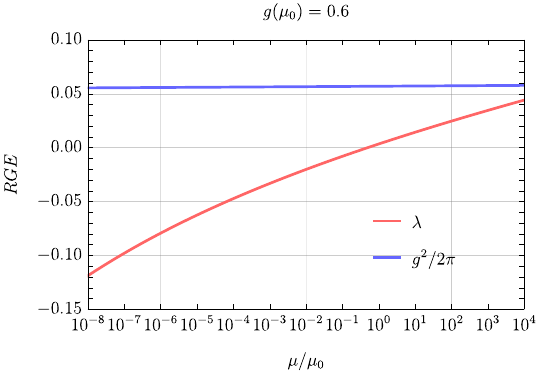}
    \hfill
\caption{The tree level scalar quartic coupling $\lambda$ as a function of the dark  gauge coupling $g $ that allows the generation of the VEV of the scalar $\langle \phi\rangle=v$, where $\lambda= \lambda \,( \mu=v)$, left panel. The effect of RGE-running on the relevant model parameters is shown in the right panel for the benchmark $g(\mu_0)=0.6$, where $\mu_0=v$ denotes the reference scale at which the couplings are defined.
}\label{fig:Runningcouplings}
\end{figure}

The couplings entering in Eq.~(\ref{eq:oneloop}) are evaluated at the renormalization scale $\mu$ through their RGE running, which we discuss in Appendix~\ref{app:RGE}. 
Following the procedure in Ref.~\cite{Coleman:1973jx}, we fix $\mu=\langle \phi_c \rangle$. 
At tree level, imposing the stationarity condition \( V_{\mathrm{eff}}^{'T=0}(\langle \phi_c \rangle)=0\) allows the quartic coupling $\lambda$ to be expressed in terms of the gauge coupling $g$ as
\begin{equation} \label{eq:quartic}
    \lambda =\frac{3 g ^4}{16 \pi^2} \left(\frac{1}{3}- \log \left(g ^2\right)\right)\,.
\end{equation}
The resulting relation between the couplings $\lambda$ and $g$ at the symmetry breaking scale $\mu=v$ is shown in Fig.~\ref{fig:Runningcouplings}.
Substituting Eq. (\ref{eq:quartic}) into the zero-temperature potential Eq. (\ref{eq:oneloop}), and neglecting $\mathcal{O} (\lambda^2)$ terms, one obtains
\begin{equation}
V_{\mathrm{eff}}^{T=0}(\phi_c)=\frac{3 g^4\phi_c ^4}{64\pi^2}\left(2\log\frac{\phi_c}{v}-\frac{1}{2}\right)\,.
\end{equation}
Expanding around the minimum of the potential as $\phi_c(x)=v+\phi(x)$, the fluctuation $\phi$ corresponds to the physical dark Higgs field.
Its mass is then obtained from the second derivative of $V_{\mathrm{eff}}^{T=0} (\phi_c)$, evaluated at its minimum, namely
\begin{equation}
m_{\phi}^2=\left.\frac{\partial^2 V_{\mathrm{eff}}^{T=0}}{\partial \phi_c^2}\right|_{\phi_c=v}
=\frac{3 g^4 v^2}{8\pi^2}\,.
\label{eq:mphi}
\end{equation}

This value is subject to cosmological constraints from Big Bang Nucleosynthesis (BBN), which we discuss in Sec.~\ref{sec:bbnbounds}.

In principle, we should have included in Eq.~(\ref{eq:Ldark}) the scale-invariant Higgs-portal interaction $ \lambda_{H\Phi}|\Phi|^2 |H|^2$. However, this term induces a mass term for $\Phi$ upon electroweak symmetry breaking, spoiling classical scale invariance of the dark sector. To obtain an estimate of how small should $\lambda_{H\Phi}$ be, we require it to be small enough that the dominant contribution to the $\Phi$ mass come from the CW potential, i.e. $\lambda_{H\Phi} v^2_\EW/2 < m_\phi^2$. Via Eq.~(\ref{eq:mphi}), this implies $\lambda_{H\Phi} \lesssim 10^{-9} (g/0.6)^4 (v/80~\text{MeV})^2$.
While very small, these values of $\lambda_{H\Phi}$ are not altered by running effects due to the particle content introduced so far, nor they will be when we further extend the model to accommodate for cosmological constraints in Sec.~\ref{sec:BBN}.

\subsection{Finite Temperature correction}
\label{sec:V_finiteT}
We now include the finite temperature corrections to the scalar potential. The relevant contribution to the effective potential is given by

\begin{equation}\label{eq:thermalpotential}
   V_{T}(\phi_c,T)= \sum_{i=\mathrm{bosons}}\frac{n_i T^4}{2\pi^2}J_b\left(x\right)\,, 
\end{equation}
where $n_i$ is the number of bosonic degrees of freedom, $x\equiv \frac{m_i^2(\phi_c)}{T^2}$ and the thermal correction is encoded in the bosonic thermal function 
\begin{equation} \label{eq:thermalfunctions}
    J_{b}(x)=\int_{0}^{\infty}\,dk \,k^2 \log{\biggl[1- \exp{\left(-\sqrt{k^2+x}\right)} \biggr]}\,,
\end{equation}
whose high-temperature expansion ($x \ll1$) reads \cite{Quiros:1999jp}:
\begin{equation}\label{bosht}
    J_b(x)=-\frac{\pi^4}{45}+\frac{\pi^2}{12}x-\frac{\pi}{6}x^{3/2}-\frac{x^2}{32}\log\frac{x}{a_b}+\mathcal{O}\biggl(x^3\log\frac{x^{3/2}}{\text{const.}}\biggr)\,,
\end{equation}
being $\log a_b\simeq5.4076$.

At high temperature, the bosonic zero modes become strongly coupled and they could spoil the perturbative expansion~\cite{Dolan:1973qd}.  
Two approaches have been proposed to tackle this problem. The first one is the so called \textit{daisy resummation} \cite{Arnold:1992rz} in four dimensions (4D). It consists in resumming the so called daisy diagrams, i.e. $N$ ring diagrams, where $N-1$ loops are attached to a main one, which are responsible for the infrared divergences. The second approach is the \textit{dimensional reduction} (DR) framework \cite{Kajantie:1995dw,Farakos:1994kx} in which the infinite tower of non-zero modes is integrated out, yielding an Effective Field Theory (EFT) of dimension three (3D). 

Ref.~\cite{Christiansen:2025xhv} recently showed that, by an optimal choice of $\mu$ (in a sense we specify in Sec.~\ref{sec:scale_dependence}) and by including the RGE effects for the couplings, the predictions of the GW parameters in the 4D one-loop Daisy resummed approach are in excellent agreement with the next-to-leading order (NLO) 3D potential in the DR approach, in particular for classically-scale-invariant potentials. This result motivates us to use the 4D theory for our calculations. 
Therefore, we consider the contribution to the potential due to the Daisy corrections, resummed according to the Arnold-Espinosa prescription \cite{Arnold:1992rz}: \begin{equation}
    V_{\mathrm{Daisy}}(\phi_c,T)=-\sum_i \frac{n_i\,T}{12 \pi} \bigg[ (m_i^2(\phi_c)+\Pi_i(T))^{3/2}-(m_i^2(\phi_c))^{3/2} \bigg ] \,,
\end{equation}
where the thermal masses $\Pi_i(T)$ are given by: 
\begin{equation}
    \Pi_{\phi,\varphi_i} = \frac{\lambda T^2}{4}+ \frac{g^2T^2}{4}, \qquad \quad \Pi_{V}=\frac{g^2T^2
    }{3}. 
\end{equation}
To sum up, the temperature-dependent effective potential is given by:
\begin{equation}\label{eq:fullpotental}
    V(\phi_c,T)\equiv V_{\mathrm{eff}}^{T=0}(\phi_c)+V_T(\phi_c,T)+V_{\mathrm{Daisy}}(\phi_c,T).
\end{equation}
%
%
The dark sector discussed in this section contains only bosonic degrees of freedom, $V$ and $\phi$. In Sec.~\ref{sec:BBN} we will add also new fermionic degrees of freedom, but we will couple them to $\Phi$ weakly enough that the potential $V(\phi_c,T)$, as built so far, remains valid to a very good approximation.

\subsection{Renormalization scale dependence}
\label{sec:scale_dependence}

Standard 4D one-loop computations of $V(\phi_c,T)$ induce a significant theoretical uncertainty in the predictions of PT quantities and associated GW, mostly due to their dependence on the renormalization scale $\mu$~\cite{Croon:2020cgk}.
This issue becomes even more significant in models with approximate scale invariance like the one of our study, because they feature large hierarchies between the physical scales relevant for the PT predictions, e.g. between the scalar VEV and the nucleation temperature.
The prescription for $\mu$ in 4D one-loop computations of $V(\phi_c,T)$ is therefore crucial. Somehow naively, one could look at the full potential in the high temperature expansion and notice that the dependence on $\mu$ is encoded in logarithms of the form  \footnote{This feature is obtained by summing the logarithmic terms of the one-loop radiative corrections at $T=0$ and the logarithmic pieces of the thermal corrections in the HT limit} $\sim \log{(T/\mu)}$.
It appears therefore natural to choose $\mu$ proportional to the temperature, where the choice of the proportionality coefficient can be guided by the different energy scales set by the thermal frequencies $\omega_n=n\pi \,T$ ($n$ is an odd integer for fermions and an even integer for bosons)~\cite{Croon:2020cgk}. However, in the typical one-loop computation, varying $\mu=\kappa T$ with $\kappa\sim\mathcal{O}(1)$ leads to a large uncertainty on the prediction of the GW parameters. 

Independently of the specific choice of renormalization scale, one can tackle the question on how to reduce the dependence on it. A way to mitigate this effect is to consider the RGE-improved effective potential (for an example in the context of supercooled PTs, see e.g.~\cite{Kierkla:2022odc}). 
Renormalization scale dependence can be seen as a hint of missing higher-order terms, and indeed it has been noted~\cite{Croon:2020cgk} that the uncertainty is largely reduced if we consider a more complete calculation that is organised in powers of couplings, keeping terms up to $\mathcal{O}(g^4)$. In the standard 4D daisy resummed framework, the calculation is incomplete up to this order since logarithms from two-loop thermal masses and vacuum diagrams (and logarithms related to field renormalisation) are missing.
Therefore a correct reduction of the scale-dependence cannot be fully achieved by just considering RGE effects at tree and one-loop level~\cite{Gould:2021oba}. 
These uncertainties significantly reduce in the 3D NLO DR framework (for examples in the context of supercooled PTs, see~\cite{Kierkla:2023von} with derivative expansion and~\cite{Kierkla:2025qyz} without it), because it allows to straightforwardly keep the relevant terms in the expansion and therefore to systematically reduce uncertainties by computing consistently higher orders (see~\cite{Chala:2025oul,Bernardo:2026whs} for the current DR analyses of PT parameters reaching the highest order).

As anticipated in Sec.~\ref{sec:V_finiteT}, in this study we follow the same strategy as in \cite{Christiansen:2025xhv}, in which the authors fixed the renormalization scale by requiring good agreement with the 3D theory in the computation of the bounce action. It should be noted that Ref.~\cite{Christiansen:2025xhv} neglected higher dimensional operators. They are expected to have a small impact on nucleation parameters (like $\Tnuc$ and $\beta/H$, see Sec.~\ref{sec:Gravsignal}) for the very strong radiatively-induced PT of our interest~\cite{Kierkla:2023von,Christiansen:2025xhv,Bernardo:2026whs}, in a nutshell because the relevant scale for those parameters is $\Tnuc \ll v$. However, higher-dimensional operators can sizeably affect the PT strength $\alpha_\GW$~\cite{Bernardo:2026whs} so that our results for $\alpha_\GW$ in Sec.~\ref{sec:Gravsignal} should only be taken as indicative. 

Coming to our calculation, ref.~\cite{Christiansen:2025xhv} obtained the best agreement between the 3D NLO potential and the 4D high-temperature potential by i) fixing the 4D renormalization scale at $\mu=\pi\,T$ and ii) including running effects.  We thus adopt the same prescription i) and ii) in all our numerical computations.

\section{Gravitational Wave Signal}
\label{sec:Gravsignal}

The GW signal produced upon a FOPT is determined by a set of macroscopic parameters, which are controlled by the tunneling rate between the false and the true vacua. These include the PT strength $\alpha_\GW$, its inverse duration $\beta$, the nucleation temperature $\Tnuc$, the reheating temperature $\Treh$ and the bubble wall velocity \(v_w\) (see \cite{Caprini:2024gyk} for a recent review). 
A reliable prediction of the PT parameters requires computing the thermal tunneling bounce action governing the bubble nucleation rate. In the following, we assume a relativistic wall velocity $v_w = 1$, which is typical of strongly supercooled PTs and provides a good approximation for the present analysis.
All the relevant quantities for the PT mentioned above are then computed from the finite temperature effective potential given in Eq. (\ref{eq:fullpotental}).

\subsection{Phase Transition Parameters} \label{sec:PTparameters}

\begin{figure}[H]
  \centering
  \begin{minipage}[t]{0.48\textwidth}
    \centering
    \includegraphics[width=\linewidth]{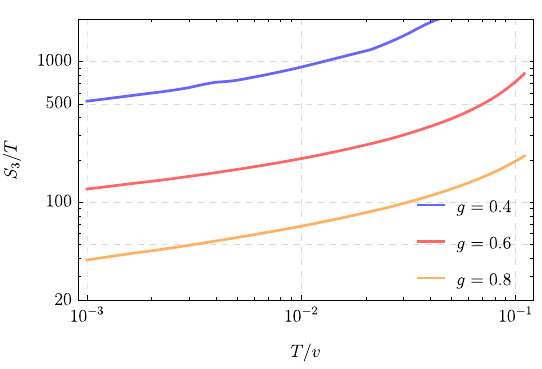}
  \end{minipage}
  \begin{minipage}[t]{0.48\textwidth}
    \centering
    \includegraphics[width=\linewidth]{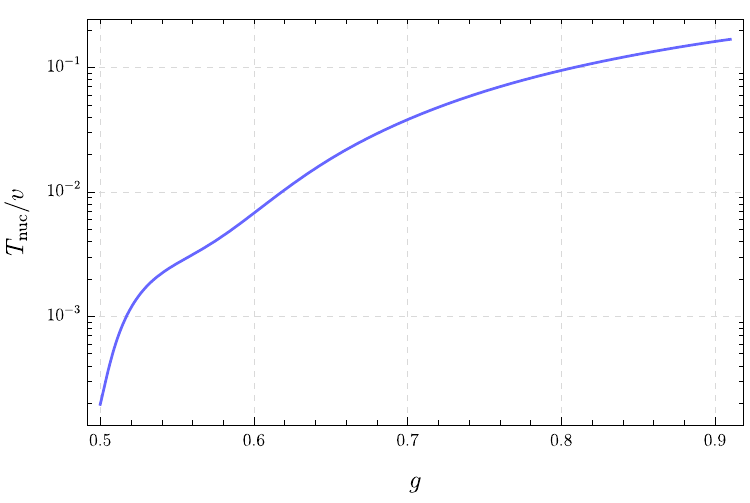}
  \end{minipage}\hfill
  \caption{ The tunneling action $S_3(T)/T$ computed from  the full one-loop potential in Eq. (\ref{eq:fullpotental}) for the benchmark values of the dark gauge coupling $g=0.4,0.6,0.8$ (left panel). The nucleation temperature $\Tnuc$ as a function of $g$, normalised to the scalar field VEV $v$ (right panel).  }
  \label{fig:Tnucleation}
\end{figure}

The tunneling rate per unit volume $\Gamma$, from the false to the true vacuum, is known to be controlled for the CW potential by thermal effects (see e.g. \cite{Levi:2022bzt}), hence it can be expressed as  
 \begin{equation} \label{eq:S3rate}
     \Gamma (T)\simeq T^4\biggl(\frac{S_3/T}{2\pi}\biggr)^{3/2}\exp{(-S_3 /T)},
\end{equation}
where $S_3$ is the 3-dimensional Euclidean action evaluated on the so-called bounce solution. We compute $S_3$ using the Mathematica package \texttt{FindBounce}~\cite{Guada:2020xnz} and verify its good agreement with the analytical expressions of~\cite{Levi:2022bzt}.
We show our results for $S_3(T)/T$ as a function of $T$ in Fig. \ref{fig:Tnucleation} (left panel) for the benchmark values of the dark gauge coupling $g=0.4,0.6,0.8$.
The value of $S_3 (T)/T$ rapidly decreases as $g$ increases, which can be understood as follows.
The cubic term encoded in the high-temperature expansion of the one-loop thermal potential Eq. (\ref{eq:thermalpotential}) lowers the potential barrier with increasing $g$, causing bubble nucleation to occur at earlier times. Hence, by increasing $g$, the PT completes at higher temperatures (see the right-hand panel of Fig. \ref{fig:Tnucleation}) and eventually takes place during radiation domination for $g\gtrsim 1$. Conversely, for small values of $g \lesssim 0.45$, the cubic term becomes ineffective, the FOPT does not complete and the Universe remains trapped in a state of eternal inflation (see \cite{Levi:2022bzt} for a detailed analysis).

The PT occurs around the \textit{nucleation temperature} $\Tnuc$, when one bubble nucleates per Hubble volume, namely $\Gamma (\Tnuc) \simeq H(\Tnuc)^4$.
The Hubble parameter is given by
\begin{equation}
    H^2(T)=\frac{1}{3 M_{\Pl}^2} \left[\rho_\text{rad} (T)+ \Delta V_0 \right]\,,
\end{equation}
where $\Delta V_0$ is the energy difference between the false and the true vacuum of the zero-temperature potential, $\rho_\text{rad}= \pi^2/30 \, g_{\ast} T^4$ the radiation energy density, with $g_{\ast}$ the number of the relativistic degrees of freedom.
We show in the right panel of  Fig. \ref{fig:Tnucleation} our computed nucleation temperature normalized to the scalar field VEV $v$, as a function of $g$. We point out that the \textit{percolation} temperature $T_p$, defined as the temperature when a connected region of the true vacuum forms over the whole Universe, was shown \cite{Athron:2022mmm} to provide a more accurate description in the case of supercooled PTs. However, as pointed out in \cite{Levi:2022bzt}, the discrepancy between $T_p$ and $\Tnuc$ is at most $20 \%$ for the values of $\beta_H$ that we obtain in our scenario, that is $\beta_H \geq \mathcal{O} (20-30)$, hence, for computational simplicity, we adopt $\Tnuc$ as our definition of the PT temperature.

\begin{figure}[H]
  \centering
\begin{minipage}[t]{0.49\textwidth}
    \centering
    \includegraphics[width=\linewidth]{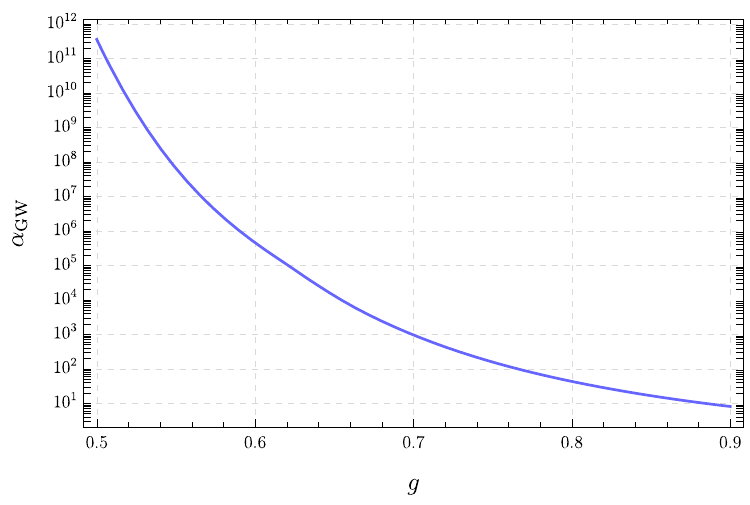}
  \end{minipage}
  \begin{minipage}[t]{0.49\textwidth}
    \centering
    \includegraphics[width=\linewidth]{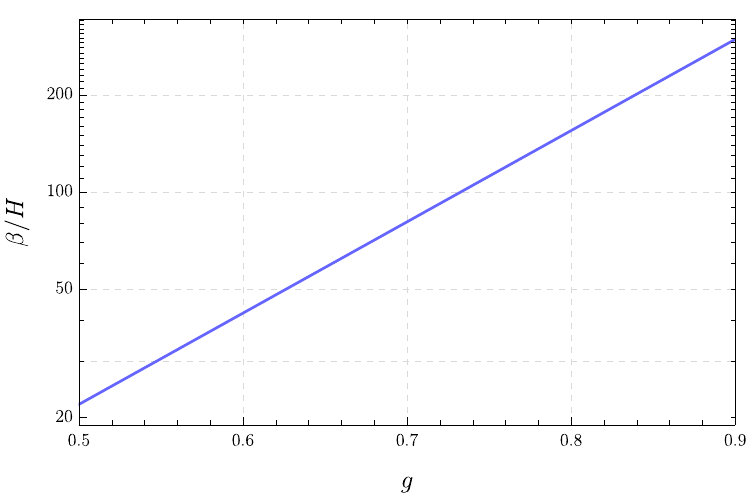}
  \end{minipage}\hfill

  \caption{The relevant GW parameters  $\alpha_\GW$ (left) and $\beta_H$ (right) as a function of the dark gauge coupling $g$. The results are shown for VEV of the scalar field $v=100$ MeV.}
  \label{fig:GWparameters}
\end{figure}

The strength of the PT, namely the energy released as bulk motion during the PT, normalized to the radiation density, is given by 
\begin{equation}
    \alpha_\GW=\frac{\Delta V_0}{\rho_{\text{rad}}(\Tnuc)}\,.
\end{equation}
The last relevant parameter that we need to compute to estimate the GW spectrum is the PT rate $\beta_H$, defined as the ratio of the inverse duration of the PT $\beta= d \log \Gamma/dt$, being $\Gamma$ the tunneling rate in Eq. (\ref{eq:S3rate}), and the Hubble parameter $H$ evaluated at the nucleation temperature, namely  
\begin{equation}
    \beta_H=T\,\frac{d(S_3/T)}{dT}\bigg|_{\Tnuc}\,.
\end{equation}
Our results for $\alpha_\GW$ and $\beta_H$ are presented as a function of $g$ in Fig. \ref{fig:GWparameters}, for a benchmark value of $v=100$ MeV.
We remind the reader that our values of $\alpha_\GW$ should only be taken as indicative, see the discussion in the end of Sec.~\ref{sec:scale_dependence}.

\subsection{Gravitational wave spectrum} \label{sec:GWtemplates}

First-order PTs produce GW signals from several sources, the most studied ones are the collision of bubble walls and the sound waves and turbulence in the plasma that follow those collisions, see e.g.~\cite{Caprini:2024hue}. We now argue why none of the GW templates above is the correct one for supercooled PTs like the one of our interest, for which actually an accurate GW template is yet to be derived, as already stressed in~\cite{Baldes:2024wuz}. 

The large values of $\alpha_\GW$ of our interest (see Fig.~\ref{fig:GWparameters}) imply that walls quickly become ultrarelativistic:  local-thermal-equilibrium hydrodynamics does not prevent the walls from becoming ultrarelativistic~\cite{Krajewski:2024gma}, and out-of-equilibrium friction does not either because it has been proven~\cite{Ekstedt:2025awx} to be limited by the mass-gain (LO) pressure of~\cite{Bodeker:2009qy}, which does not stop the walls at large $\alpha_\GW$.
Ultrarelativistic walls imply that sound waves and turbulence are a subdominant source of GW.

If walls run away, i.e. if they keep accelerating until they collide, then they keep most of their energy without transferring it to the plasma. The GW they emit are then described by the ``bulk flow'' model~\cite{Konstandin:2017sat,Jinno:2017fby}, recently improved into a ``dissipative bulk flow''~\cite{Lewicki:2025hxg} that accounts both for cosmic expansion at GW production and for the additional energy dissipation of scalar field shells after the collisions, relevant for very slow PT (i.e. small $\beta/H$).
However, radiation of one (NLO)~\cite{Bodeker:2017cim} and more (NLL)~\cite{Gouttenoire:2021kjv,Azatov:2023xem} gauge bosons off incoming particles induces a pressure that grows linearly with the boost-factor of bubble walls. In PT models like ours, that radiation pressure implies that walls do not run away, but reach a terminal (ultrarelativistic) velocity before colliding.
In this regime most of the energy is transferred from the walls to the plasma, and we are not aware of any calculation of the associated GW spectrum.
Moreover, the shells that inevitably form at ultrarelativistic walls~\cite{Baldes:2024wuz} could well have a further impact on the GW spectrum, whose exploration has not started yet.

The GW template for the PT of our interest, which is unknown, would arguably be closer to bulk-flow-like templates than to sound waves and turbulence ones. The core reason is that, in PTs with ultrarelativistic non-runaway walls, most of the energy stays in a very thin region close to bubble walls, like it happens when walls runaway. We refer the reader to~\cite{Baldes:2024wuz} for an extended discussion supporting this expectation. We therefore use a GW template from the collision of runaway bubble walls, and explain our specific choice next.

\medskip 

A novel fit of the PTA GW signal in terms of the PT parameters ($\beta/H$, $\alpha_\GW$,...) goes well beyond the purpose of this paper, so we rely on existing fits.
In light of the discussion above, we look for fits of the PTA signal with GW templates from runaway colliding walls. Within that category we look for the most conservative fits, i.e. the ones that result in the wider areas for the PT parameters, because of all the uncertainties and ``known unknowns'' that we discussed above\footnote{As well as because of those that we did not discuss, like for example non-sphericities of bubbles~\cite{Bian:2026xdm}}.
These criteria motivate us to choose the fit of the PTA signal of the NANOGrav collaboration itself~\cite{NANOGrav:2023hvm}, which relies on the ``envelope'' GW template~\cite{Jinno:2016vai} (a precursor of bulk flow) and which is particularly conservative because it allows for the possibility that part of the observed signal comes from SMBH binaries.

\begin{figure}[H]
        \centering
        \includegraphics[width = 0.8\textwidth]{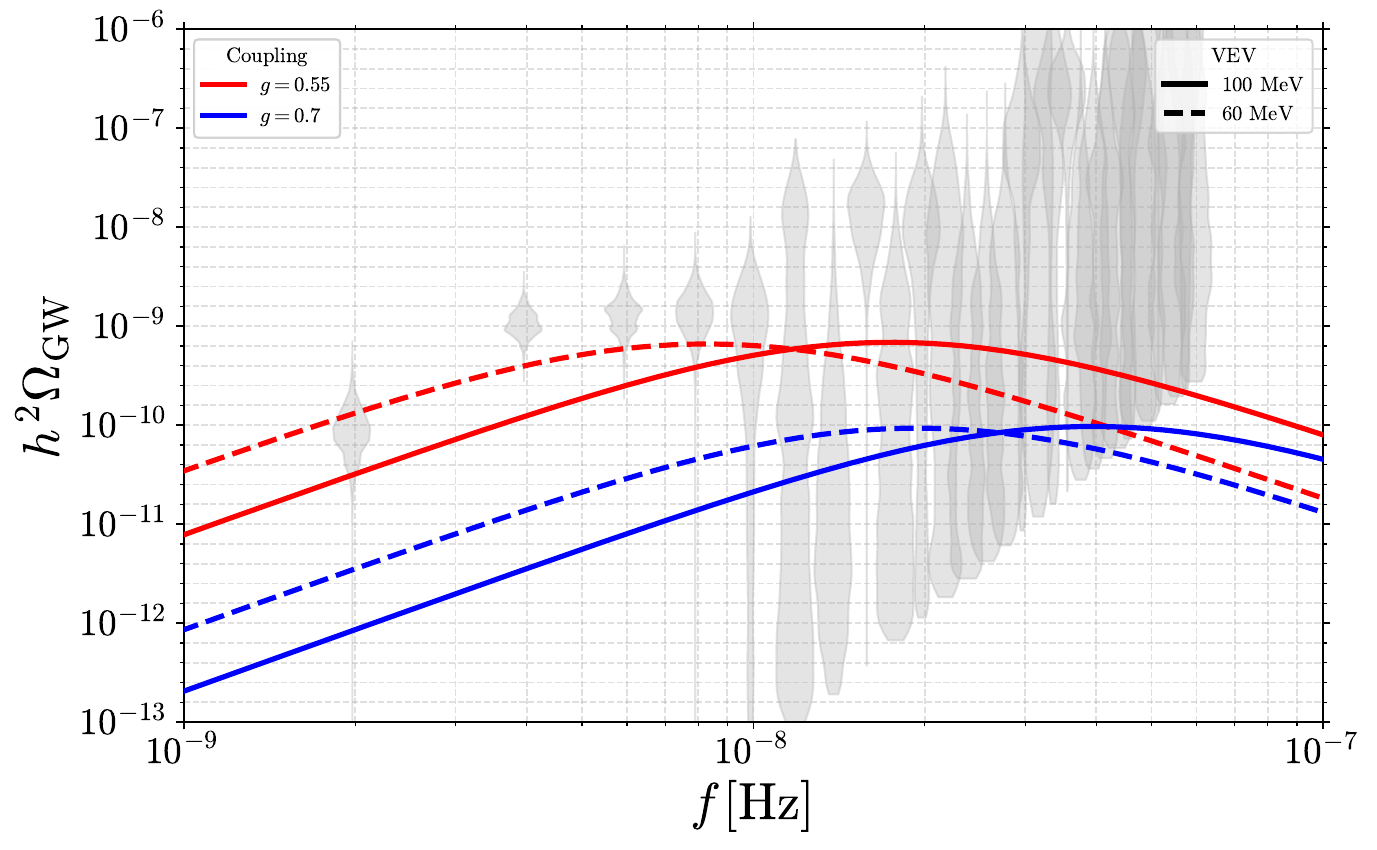}
        \caption{ Predicted gravitational wave spectrum from bubble wall collisions using the envelope template~\cite{Jinno:2016vai} adopted by the NANOGrav collaboration~\cite{NANOGrav:2023hvm},
        for benchmark values of the dark gauge coupling $g=0.55$ (red) and $g=0.7$ (blue) and scalar field VEV $v=100$ MeV (solid) and  $v=60$ MeV (dashed). The gray regions correspond to the NANOGrav's violin plot \cite{NANOGrav:2023gor}. See Fig.~\ref{fig:gwtemplates} in App.~\ref{app:GWtemplates} for a comparison with other GW templates and Sec.~\ref{sec:GWtemplates} for a critical discussion of the GW templates appropriate for our type of PT.}
                \label{fig:gwspectrum}

\end{figure}

We thus show in Fig.~\ref{fig:gwspectrum} the GW spectra computed using the bubble wall contribution in the envelope approximation~\cite{Jinno:2016vai}, for benchmark values of the gauge coupling $g=0.55, 0.7$ and of the scalar field VEV $v=100$ MeV and $v=60$ MeV.
We compare those spectra with the "violins'' from the NANOGrav 15-year data set~\cite{NANOGrav:2023gor}, representing the observed signal. 
More details about the envelope GW template are provided in App.~\ref{app:GWtemplates}. Because of the many uncertainties and unknowns,  there we also illustrate the templates for dissipative bulk flow and sound waves plus turbulence, and compare them among each other and the envelope one.

\section{Signals and constraints}
\label{sec:constraints}

After having established the parameter space for successful baryogenesis in Sec.~\ref{sec:baryonasymmetry} and computed the GW spectrum in Sec.~\ref{sec:Gravsignal}, in this Section we identify the parameter space where our model reproduces the observed BAU and at the same time the PTA nHz GW, while being allowed by all cosmological, astrophysical and laboratory constraints that we are aware of.

\begin{figure}[H]
        \centering
        \includegraphics[width = 0.7\textwidth]{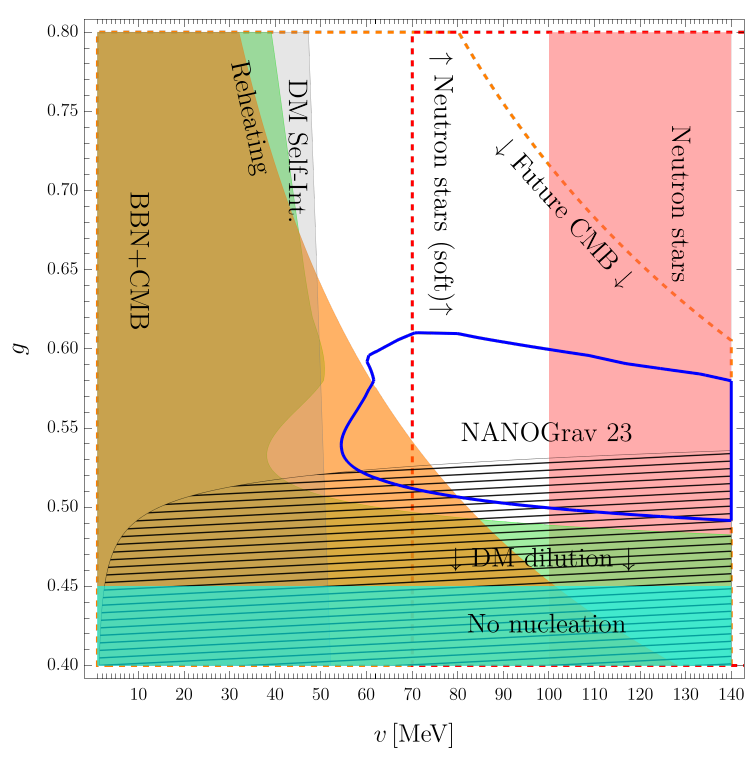}
        \caption{The baryon asymmetry of the universe is successfully reproduced in the entire $(g,v)$ plane upon suitable choices of $\Delta m$ and $\delta m$, see Sec.~\ref{sec:baryonasymmetry}.
        The blue contour encloses the parameter space compatible with the NANOGrav 15-year nHz GW signal, according to its conservative fit in~\cite{NANOGrav:2023hvm} and to our calculation of the PT parameters of Secs~\ref{sec:potential} and~\ref{sec:PTparameters}.
        Colored regions are excluded by insufficient reheating temperature after the PT $T_\text{reh} < 6.5$~MeV (green, Sec. \ref{sec:darkbaryon}), absence of nucleation (cyan, Sec.~\ref{sec:PTparameters}), dilution of the DM relic abundance due to $\alpha_\GW > 10^9$ (dashed black, Sec. \ref{sec:DMabundance}), DM self-interactions in galaxy clusters $\sigma^\text{self}/m_\chi > 0.35$~cm$^2/$gr (gray, Sec. \ref{sec:SIDM}), neutron stars observations (red, Sec. \ref{sec:Neutronstars}), BBN+CMB bounds $m_\phi > 4$~MeV (orange, Sec.~\ref{sec:bbnbounds}).
        The dashed orange line indicates the sensitivity achievable with future CMB experiments $m_\phi \simeq 10$~MeV, the dashed red one the more aggressive NS exclusion using another equation of state.}
            \label{fig:parameterspace}

\end{figure}

\subsection{Baryogenesis and PTA parameter space}

In Fig.~\ref{fig:parameterspace} we display as a white area the region in the $(g,v)$ plane that is compatible with the baryogenesis mechanism discussed in Sec.~\ref{sec:baryonasymmetry}.
We stress that the viability of the baryogenesis mechanism depends on the interplay between additional parameters, such as the neutron-DM mass splitting $\Delta m$ and the mixing $\delta m$ between the two states, discussed in Sec.~\ref{sec:baryonasymmetry}. 
However, for each point of $(g,v)$ consistent with all the constraints, there is a suitable choice of $(\delta m,\Delta m)$ that reproduces the correct BAU, see Sec.~\ref{sec:baryonasymmetry} and the accelerator constraints on $(\delta m,\Delta m)$ in Sec.~\ref{sec:accelerator}.

The blue contour encloses the region where the parameters $g$ and $v$, via our calculations in Sec.~\ref{sec:PTparameters}, imply values of $\beta/H$, $T_\text{nuc}$ and $\alpha_\GW$ that fall within the $95\%$ C.L. NANOGrav best-fit region~\cite{NANOGrav:2023hvm}.
Note that any value of the PT strength $\alpha_\GW \gtrsim 0.1$ is favored by the best-fit. Therefore the uncertainties in $\alpha_\GW$ induced by our procedure (see Sec.~\ref{sec:scale_dependence}) do not have an impact on the blue contour, because we obtain anyway $\alpha_\GW \gg 1$ (see Fig.~\ref{fig:GWparameters}).
As anticipated in Sec.~\ref{sec:scale_dependence}, we have used the NANOGrav best-fit because it is the most conservative one that we are aware of, as it considers the combined contribution of SMBH binaries and a FOPT (where the GW are modelled using the envelope approximation of~\cite{Jinno:2016vai}).

The dynamics of the PT excludes the region in cyan, where bubble nucleation is not achieved (see discussion in Sec.~\ref{sec:PTparameters}), and in green, where the reheating temperature after the PT is too small (see discussion in Sec.~\ref{sec:darkbaryon}). We now discuss the other constraints displayed in Fig.~\ref{fig:parameterspace}.

\subsection{Accelerators}
\label{sec:accelerator}
Searches for monojet+MET with 140~fb$^{-1}$ at the LHC~\cite{ATLAS:2024vqf} have been recasted in~\cite{Hiller:2026osz} on the dimension-6 operator $\mathcal{O}_6 = udd\chi/\Lambda_6^2$, finding limits $\Lambda_6 \gtrsim \Lambda_6^\text{LHC} = 9$~TeV for $M_\chi = 1$~GeV. We first note that those limits also apply to the values of $M_\chi \simeq m_n$ of our interest, which are not displayed in~\cite{Hiller:2026osz}.
We estimate the impact of the same searches on the operator of our interest of Eq.~(\ref{eq:effectiveop7}), $\mathcal{O}_7 = udd\chi\phi/\Lambda_7^3$, as follows. 
The cross section for producing monojet+MET scale, due to $\mathcal{O}_6$ scales as $\sigma_6 \propto \hat{s}/\Lambda_6^4$, where $\hat{s}$ is the scale relevant for the process which we assume to be of the order of the cuts used in~\cite{ATLAS:2024vqf}, so $\hat{s} \simeq$~TeV$^2$.
Coming now to $\mathcal{O}_7$, the initial parton state is the same as in $\mathcal{O}_6$, and the final state with respect to the one due to $\mathcal{O}_6$ contains an additional $\phi$, which is extremely long-lived on collider scales and thus contributes to MET.
We can therefore estimate $ \sigma_7 \propto \hat{s}^2/\Lambda_7^6 \times 1/(16\pi^2)$,
with the same proportionality coefficient as $\sigma_6$ and where $1/(16\pi^2)$ comes from the extra $\phi$ in the final state. We thus obtain the limit
\begin{equation}
\Lambda_7 > \Lambda_7^\LHC \simeq \Lambda_6^\LHC \Big(\frac{\hat{s}}{4 \pi \Lambda_6^\LHC}\Big)^\frac{1}{3}
\simeq 1.9~\text{TeV}\,,
\label{eq:Lambda7_LHC}
\end{equation}
which is roughly still within the validity of the EFT used in~\cite{Hiller:2026osz} to recast~\cite{ATLAS:2024vqf}.
Eq.~(\ref{eq:Lambda7_LHC}) corresponds to
\begin{equation}
\delta m \lesssim 1.5 \times 10^{-10}~\text{MeV} \times \frac{v}{100~\text{MeV}}
\,.
\end{equation}
This upper limit is independent of $\Delta m$ and dominates over the one from $n \to \chi \gamma$ for $\Delta m \lesssim 0.2$~MeV, see Fig.~\ref{fig:Baryogenesis_ParameterSpace}.
Note that these limits did not exist at the time this model was first proposed~\cite{Cline:2018ami,Bringmann:2018sbs}.

Limits from hyperon decays~\cite{Alonso-Alvarez:2021oaj}, as well as from decays of $B$ and $D$ mesons~\cite{Hiller:2026osz}, are much weaker than the one in Eq.~(\ref{eq:Lambda7_LHC}) unless $\mathcal{O}_7$ also contains strange, or heavier, quarks. Even if $\mathcal{O}_7$ contained these heavier quarks, the rescaling of the $\Lambda_6$ limits would make the $\Lambda_7$ ones much weaker, because now the additional energy scale suppressing the limit would be $\hat{s} \simeq m_\text{meson}^2$.\footnote{The same applies to limits from supernovae~\cite{Alonso-Alvarez:2021oaj}, where $\hat{s} \simeq T^2_\SN$ and $T_\SN = $ tens of MeV.} 

Given that the values of $\Lambda_7$ needed by the baryogenesis mechanism are between 2 and 4 TeV (see Fig.~\ref{fig:Baryogenesis_ParameterSpace}), one should worry about possible additional constraints on EW invariant UV completions of the operator $\mathcal{O}_7$ of Eq.~(\ref{eq:effectiveop7}).
We consider its UV completion of~\cite{Cline:2018ami}, where one adds to the SM two scalars $\Phi_1$ and $\Phi_2$, both triplets under $SU(3)$ with hypercharge $1/3$ and baryon number $-2/3$, and where only $\Phi_1$ is charged under $U(1)_\D$. The resulting renormalizable Lagrangian then is
\begin{equation}
\mathcal{L}_\UV =
\lambda_1 \, \bar{d} P_L \chi \, \Phi_1 + \lambda_2 \, \overline{u^c} P_R d \, \Phi_2 + \mu \,\phi\, \Phi_1 \Phi_2^\dagger + \text{h.c.}\,,
\label{eq:EWinvariant}
\end{equation}
leading to
\begin{equation}
\Lambda_7^3 = \frac{m_{\Phi_1}^2 m_{\Phi_2}^2}{\lambda_1 \lambda_2 \mu}\,,
\end{equation}
where $m_{\Phi_{1,2}}$ is the mass of $\Phi_{1,2}$ and $u,d$ are the SM up and down quarks.
Eq.~(\ref{eq:EWinvariant}) gives rise to other signals not discussed so far, some of which were missed in~\cite{Cline:2018ami}. Direct searches for doubly-produced $\Phi_{1,2}$ are analogous to squark searches and constrain $m_{\Phi_{1,2}}\gtrsim 800$~GeV~\cite{CMS:2020zti}.
Limits from singly-produced scalar triplet diquarks apply to $\Phi_2$ but put limits in the same ballpark for $O(1)$ couplings $\lambda_2$, see e.g.~\cite{Pascual-Dias:2020hxo}\footnote{
Strictly speaking, we are not aware of LHC recasts to our exact $u_R^c d_R \Phi_2$ structure, but we expect limits will be in the same ballpark of those from the recast in~\cite{CMS:2020zti}, which are anyway much weaker than the EFT limits unless $\lambda_{1,2} \ll 0.1$.}.
The vertex $\bar{d} P_L \chi \, \Phi_1$ gives rise to so-called $t$-channel DM production, see e.g.~\cite{Papucci:2014iwa}. It is strongly constrained by searches for multijet plus missing transverse energy (MET) at CMS~\cite{CMS:2021far} (see the ``fermion portal'' interpretation), $m_{\Phi_1}/\lambda_1 \gtrsim 1.5~\text{TeV}$, which is consistent with the earlier ATLAS ``scalar-colour-charged'' limit in~\cite{ATLAS:2019wdu}. 
Coming to the dimension-6 operators induced by Eq.~(\ref{eq:EWinvariant}), we assume that flavor-violating couplings are negligible, otherwise they would dominate the exclusions~\cite{Giudice:2011ak}.
We finally turn to the interaction
$(\lambda_2^2/2m_{\Phi_2}^2) (\bar{u}_R \gamma_\mu u_R) (\bar{d}_R \gamma_\mu d_R) $,
which is strongly constrained by searches for BSM in dijet angular distributions at CMS~\cite{CMS:2018ucw} (we use the limit on $\Lambda_\text{RR}^+$), $m_{\Phi_2}/\lambda_2 \gtrsim 3.6~\text{TeV}$.
Multijet+MET and dijet limits are then stronger than those coming from the direct production of the scalars discussed above, unless $\lambda_1 \lambda_2 \lesssim 0.1$.
We can combine the above monojet and dijet limits as a limit on $\Lambda_7$ as
\begin{equation}
\Lambda_7 \gtrsim 1.9~\text{TeV}\times \left(\frac{\lambda_1 \lambda_2}{0.1} \frac{400~\text{GeV}}{\mu}\right)^\frac{1}{3}\,.
\end{equation}
We then learn that limits coming from the UV completion of Eq.~(\ref{eq:EWinvariant}) are weaker than those on the dimension-7 operator of Eq.~(\ref{eq:Lambda7_LHC}), displayed in Fig.~\ref{fig:Baryogenesis_ParameterSpace}, if $\mu/(\lambda_1\lambda_2) \gtrsim 4$~TeV.

\subsection{Dark Matter relic abundance}\label{sec:DMabundance}
In the present work we assume the DM $\chi$ to carry a primordial asymmetry, such that its present-day relic abundance is set by its asymmetric component. In order for this scenario to be viable, the symmetric DM component must efficiently annihilate in the early Universe. In our setup DM annihilations dominantly occur through
$\chi \bar{\chi} \rightarrow V V$.
The thermally averaged annihilation cross section in the non-relativistic limit is given by \cite{Tulin:2012wi}
\begin{equation}
    \langle \sigma v \rangle 
    \simeq \frac{g^4}{16 \pi m_\chi^2}\,.
\end{equation}
Efficient depletion of the symmetric component requires
$\langle \sigma v \rangle  \gtrsim  5 \times 10^{-26} \; {\rm cm}^3/{\rm s}$,
where the numerical value corresponds to the standard freeze-out estimate for symmetric Dirac DM~\cite{Saikawa:2020swg}. This translates into the lower bound $g  \gtrsim  7 \times 10^{-3}$ for $m_\chi \approx m_n$.
Hence, for the benchmark values of $g$ considered in this work, this condition is safely satisfied, ensuring that the symmetric DM component is efficiently erased prior to freeze-out.

In addition, the entropy injection associated with the PT can dilute any pre-existing asymmetry. The corresponding dilution factor is $\Delta \simeq (T_{\rm reh}/T_{\rm nuc})^3$. Assuming instantaneous reheating, one finds $T_{\rm reh} \simeq (1+\alpha_{\rm GW})^{1/4} T_{\rm nuc}$~\cite{Ellis:2018mja}, so that $\Delta \simeq (1+\alpha_{\rm GW})^{3/4}$. Requiring that the observed baryon asymmetry $Y_B^{\rm obs}\simeq 8.6\times 10^{-11}$ is not overly diluted implies $\Delta \lesssim Y_B^{\rm initial}/Y_B^{\rm obs}$. Since in our setup the baryon asymmetry is transferred from the dark sector rather than generated within it, we remain agnostic about the origin of its initial value $Y_B^{\rm initial}$ and conservatively set an upper bound on $\alpha_{\rm GW} \leq 10^9 $, implying $Y_B^{\rm initial} \leq 4.8 \times 10^{-4}$.
This upper limit is to be taken only as indicative a fortiori because of the uncertainties in our calculation of $\alpha_\GW$ discussed in Sec.~\ref{sec:scale_dependence}.
A detailed discussion of the thermal history, together with an explicit realization of this reheating scenario, is presented in Sec.~\ref{sec:BBN}.

\subsection{Dark matter self-interactions}\label{sec:SIDM}
DM self-interactions, if large enough, affect the spatial shape of DM-supported structures ranging from dwarf spheroidal galaxies to galaxy clusters, see~\cite{Tulin:2017ara} for a review. They are therefore tested by observations of these systems.
Some observations seemingly contradict the results of numerical simulations of collisionless cold DM, especially at the small scales of dwarf spheroidal galaxies. A subset of these discrepancies could be solved by DM self-interactions $\sigma^\text{self}/m_\chi \approx 1$~cm$^2/$gr.
However, most of the claimed tensions became less solid or disappear once baryonic and/or systematics effects are taken into account, see~\cite{Sales:2022ich} for a review.\footnote{More recent works find that cold DM is consistent also with the diversity~\cite{Cruz:2025wfz} and plane-of-satellite~\cite{Sawala:2022xom} problems.}
In this work we take the conservative point of view that these observations impose upper limits on DM self-interactions, keeping in mind that more work is needed to assess if they may end up preferring $\sigma^\text{self}/m_\chi$ around our exclusion boundaries.

The strongest upper limits on DM self-interactions come from galaxy clusters. Although the relatively mild impact of baryonic feedback on these shapes adds to the robustness of these limits, their extraction still relies on observational and modeling systematics.
Therefore here we adopt the upper limit $\sigma^\text{self}/m_\chi < 0.35$~cm$^2$/gr of~\cite{Sagunski:2020spe}, despite slightly stronger limits exist in the literature (e.g. 0.13~cm$^2$/gr~\cite{Andrade:2020lqq} and 0.19~cm$^2$/gr~\cite{Eckert:2022qia}), and we point the reader to~\cite{Adhikari:2022sbh} for a recent review of the astrophysics involved and for a critical summary of existing limits.

\begin{figure}[t]
        \centering
        \includegraphics[width = 0.6\textwidth]{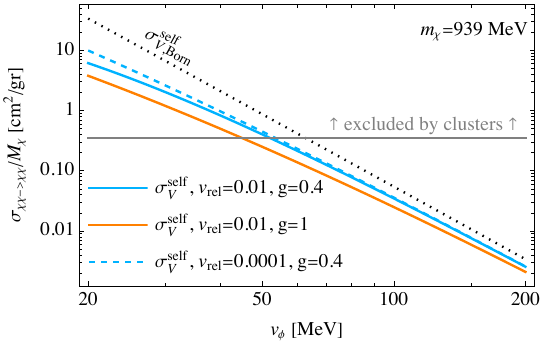}
        \caption{Self-interacting ``viscosity'' cross sections divided by DM mass as a function of the VEV $v$, for the values of DM relative velocity $v_\text{rel}$ and dark $U(1)$ coupling $g$ specified in the caption (repulsive interaction). The dotted black line reports the Born limit, valid for $\alpha \,m_\chi/m_V \ll 1$, which is independent of $g$ (and of $v_\text{rel}$ in the $v$ region displayed). Values of the cross sections for $v_\text{rel} \approx 10^{-2}$, that are larger than $\approx 0.35$~cm$^2$/gr, are excluded by observations of galaxy clusters~\cite{Sagunski:2020spe}.}
            \label{fig:self-interaction}

\end{figure}

We now discuss the prediction for $\sigma^\text{self}$ in the model of our interest.
Small-angle forward scatterings are negligible in heat conduction and thus in affecting the shape of DM halos, therefore the commonly-used prescription is to use a ``transfer'' cross section suppressing forward scattering, or a ``viscosity'' cross-section suppressing both forward and backward scattering.
The latter is more appropriate in the case of our interest of identical DM particles, and it is defined as~\cite{PhysRevA.60.2118} $\sigma_V \equiv \int d\Omega \, \sin^2\theta \,d\sigma_\text{self}/d\Omega$. 
In the parameter space of our interest, $\alpha \,m_\chi/m_V \sim O(1)$, where $\alpha=g^2/(4\pi)$, so that the Bohr radius of DM bound states is of the same order of the interaction range and long-range effects start to become important. 
The self-interacting cross section has been derived analytically in~\cite{Tulin:2013teo}, approximating the Yukawa with the Hulth\'en potential and in the limit $\epsilon_v \equiv m_\chi v_\text{rel}/(\kappa \, m_V) \ll 1$ which holds in the parameter space of our interest, where $\kappa \approx 1.6$ is a dimensionless constant. We start from that result and adjust its prefactor to take into account that we consider identical DM particles and the viscosity (as opposed to transfer) cross section, so we use
\begin{equation}
\sigma_V^\text{self}
= \frac{16 \pi}{3} \frac{1}{m_\chi^2 v_\text{rel}^2} \sin^2\delta_0,
\quad
\delta_0 = \arg\Big(\frac{i\Gamma(i\epsilon_v)}{\Gamma(\lambda_+)\Gamma(\lambda_-)}\Big),
\quad
\lambda_{\pm} = 1 + i\frac{\epsilon_v}{2} \pm i\sqrt{ \frac{\alpha \, m_\chi}{\kappa \,m_V}+\Big(\frac{\epsilon_v}{2}\Big)^2}\,,
\label{eq:sigma_self}
\end{equation}
where the $ + (-)$ sign refers to repulsive (attractive) self-interactions. As in our scenario the DM is assumed to be asymmetric,  only $\chi \chi $ scatterings are present, which correspond to the repulsive case.
  Eq.~(\ref{eq:sigma_self}) is displayed in Fig.~\ref{fig:self-interaction} for two values of the DM relative velocity relevant for clusters ($v_\text{rel} \simeq 10^{-2}$) and for dwarf spheroidal galaxies ($v_\text{rel} \simeq 10^{-4}$). At large $v$ the cross sections of Eq.~(\ref{eq:sigma_self}) correctly tend to the tree-level ``Born'' (i.e. $\alpha \,m_\chi/m_V \ll 1$) viscosity cross section of~\cite{Girmohanta:2022dog}, which we also display in Fig.~\ref{fig:self-interaction}.

\subsection{Neutron star constraints}\label{sec:Neutronstars}
Bounds on the VEV of the PT can be derived from the mass-radius relation of neutron stars (NSs).
For a given equation of state (EoS) of NSs, one can solve numerically the Tolman-Oppenheimer-Volkoff (TOV) equations of general relativistic hydrostatic structures~\cite{Tang:2018eln, Oppenheimer:1939ne} to obtain the mass-radius curves of NS.
An observational constraint on EoS, both in the absence and in presence of BSM, is that they allow for the existence of the heaviest NS observed, which have mass $M_\NS \simeq 2 M_\odot$~\cite{Demorest:2010bx}.
%

A new state $\chi$ that carries baryon number and has a mass close to that of the neutron, like our DM, can alter the equation of state (EoS) of NSs and lower the largest possible NS mass. This was first studied in~\cite{McKeen:2018xwc}, that found that NS observations would completely exclude our framework unless $\chi$ has repulsive self-interactions. As seen in Sec.~\ref{sec:SIDM},  our model predicts exactly these repulsive self-interactions, due to exchanges of $V$. NS constraints then turn into limits on the ratio $m_V/g$, first determined in~\cite{Cline:2018ami} and more recently reassessed in~\cite{Zhou:2023ndi}. As they carry some dependence on the NS EoS, we conservatively use the limit
\begin{equation}
v = \frac{m_V}{g} \lesssim 100\;\text{MeV},
\label{eq:NSlimit}
\end{equation}
required by imposing that NS as heavy as $2 M_\odot$ can be realised, and using a `stiff' equation of state for the NS. Using the alternative `DD2' equation of state~\cite{Zhou:2023ndi} would result in a stronger limit, $v \lesssim 70$~MeV, and underlines the importance of NS as probes of this mechanism.
The very recent observation of a NS with $M_\NS = (2.35 \pm 0.11) M_\odot$~\cite{Romani:2022jhd,Romani:2025ytn}, if confirmed, would induce a reassessment of the standard NS EoS, so that also the limits of~\cite{Cline:2018ami,Zhou:2023ndi} should be re-evaluated in light of that progress.

\subsection{Limits on the kinetic mixing of the dark photon}
\label{sec:epsilon}
The direct detection cross section for DM-proton scattering, which we define as $\sigma_{p\chi} = \sigma_{Z\chi}/Z^2$ where $\sigma_{Z\chi}$ is the DM cross-section for scattering off a nucleus with atomic number $Z$ and mass $M_Z \gg m_\chi$, reads (see e.g.~\cite{Pospelov:2007mp})
\begin{equation}
\sigma_{p\chi} = 4\,\epsilon^2 g^2 \alpha_\text{em} \frac{m_\chi^2}{m_V^4}
\simeq 1.3\times 10^{-39}~\text{cm}^2 \, g^2 \left(\frac{\epsilon}{10^{-7}}\right)^2 \left(\frac{50~\text{MeV}}{m_V}\right)^4\,.
\end{equation}
The strongest limit on $\sigma_{p\chi}$ at $m_\chi = 939$~MeV comes from CRESST and reads $\sigma_{p\chi} \lesssim 8 \times 10^{-39}~\text{cm}^2$~\cite{CRESST:2019jnq}, which imposes $\epsilon g \lesssim 2.5\times 10^{-7} \times (m_V/50~\text{MeV})^2$.\footnote{Limits from DarkSide-50~\cite{DarkSide:2022dhx} would be stronger, but they are based on the Migdal effect in liquids, whose theoretical calculation is in conflict with experiments~\cite{Xu:2023wev} (experiments instead agree with theory in gas~\cite{Yi:2026fmf}). SuperCDMS limits are also based on the Migdal effect, but on crystals for which the prediction is arguably more reliable, see e.g.~\cite{Esposito:2025iry}.}

Let us now discuss searches for the dark photon for the masses $30~\text{MeV} \lesssim m_V \lesssim 80$~MeV of our interest. Given the particle content introduced so far, $V$ decays to $e^+e^-$ and astrophysical and cosmological constraints then impose $\epsilon \lesssim \text{few}\times 10^{-13}$~\cite{Caputo:2025avc}. 
We anticipate that, to accommodate cosmological constraints, in Sec.~\ref{sec:BBN} we will introduce new particle content that makes $V$ decay invisibly. This results in missing energy at collider and beam dump experiments, which implies that accelerator constraints~\cite{Graham:2021ggy} are now weaker than those derived just above from DM direct detection.
Astrophysical limits are also weakened with respect to the non-invisible-decay case of~\cite{Caputo:2025avc}, but supernova (SN) cooling still excludes further parameter space~\cite{Chang:2018rso}. Combined with direct detection and for the $m_V$ and $g$ values of our interest, accelerators and SN cooling leave the following viable regions:
\begin{equation}
\text{few}~\times 10^{-7} \gtrsim \epsilon \gtrsim \text{few}~\times 10^{-8}, \qquad \epsilon < 10^{-9}.
\end{equation}
Note that the region $\text{few}~\times 10^{-8} \gtrsim \epsilon > 10^{-9}$, excluded by SN, may need to be reconsidered once we enlarge the model with fermions in Sec.~\ref{sec:BBN} to respect BBN limits. As that $\epsilon$ region plays no particular role in our model, that can comfortably live at $\epsilon < 10^{-9}$, we do not explore the reassessment of SN limits in this work.

\subsection{Big Bang Nucleosynthesis}
\label{sec:bbnbounds}

Additional constraints on the parameter space arise from BBN. 
In the minimal conformal dark $U(1)_\D$ scenario discussed in Sec.~\ref{sec:potential}, 
the dark scalar $\phi$ has a mass of order a few MeV and is the lightest state in the dark sector. 
As a consequence, $\phi$ must decay before BBN, since late decays could inject entropy into the SM plasma and modify the expansion history, thereby affecting the effective number of relativistic species $N_{\rm eff}$.

For our analysis, we adopt the bounds derived in Ref.~\cite{Sabti:2021reh} for a neutrinophilic scalar. In this scenario, requiring that the scalar decays sufficiently early to avoid spoiling BBN implies $m_\phi \gtrsim 1.3\,\mathrm{MeV}$, while including constraints from the CMB, in particular on $N_{\rm eff}$, strengthens the bound to $m_\phi \gtrsim 4\,\mathrm{MeV}$. Future CMB experiments will potentially test up to $m_\phi \simeq 10\,\mathrm{MeV}$~\cite{Sabti:2019mhn}.

In our minimal setup, the only coupling between the dark and visible sectors arises from the kinetic mixing $\epsilon$ of the dark photon. 
This induces $\phi$ decays into SM fermions at loop level. However, the corresponding decay width is highly suppressed, even for the largest allowed values $\epsilon = \text{few}\times 10^{-7}$. 
Using the expression from~\cite{Ferber:2023iso} for the $\phi$ decay width into electrons, the only kinematically allowed channel, we obtain
\begin{equation}
   \Gamma_{\phi\to e^{+}e^{-}} = \frac{27 \, e^4 \epsilon^4 \, g^2 \, m_\phi}{512\pi^5} \frac{m_e^2}{m_{V}^2} \simeq 2.6\times 10^{-41}~\text{GeV} \, \left(\frac{g}{0.6}\right)^2 \left(\frac{\epsilon}{10^{-7}}\right)^4 \left(\frac{100~\text{MeV}}{v}\right)\,.
   \label{eq:width_scalar}
\end{equation}
This width is way too small not only to satisfy BBN limits, but also to ensure efficient reheating of the SM after the PT. Our model therefore needs to be extended, we present one such extension in~Sec.~\ref{sec:BBN}.

\section{Thermal history of the dark sector}
\label{sec:BBN}

\subsection{Dark sector model}

\begin{table}[t]
\centering
\begin{tabular}{ c | c c c c c }
\toprule
& $\Phi$ & $N$ & $\chi$  & $\Psi_L$ & $\Psi_R$\\ 
\hline
$\rm U(1)_B$  & 0 & 0  & 1 & 0 & 0 \\
$\rm U(1)_\D$  & 1 & 0 & 1 & 1 & 1 \\
\bottomrule
\end{tabular}
\caption{Particle content of the dark sector model specified by the Lagrangian Eq. (\ref{eq:lag}) and charges of the fields  under the baryon and the dark sector numbers.}
\label{tab:content}
\end{table}

Motivated by the cosmological requirements discussed in Sec.~\ref{sec:bbnbounds}, we extend the fermionic sector of the minimal conformal $U(1)_\D$ model to induce sufficiently early decays of the dark scalar $\phi$ and efficient reheating of the visible sector. 

Adopting the dark sector structure of \cite{Abdullahi:2020nyr}, we introduce $n \geq 2$ generations of sterile Majorana fermions $N_R$ and $d$ generations of vector-like fermions $\Psi$, the latter charged under the dark $U(1)_\D$, see Tab. \ref{tab:content} for a summary of the field content of the model and of the field representations. The relevant interaction and mass terms are given by 

\begin{equation}
\mathcal{L} \supset
-\,y_\nu\, \overline{L} H N_R
-\,y_L\,\Phi\,  \overline{\Psi}_L N_R -\,y_R\,\Phi\,\overline{\Psi}_R^c N_R
-\,\frac{1}{2} \overline{N_R} \, M_N\, N^c_R
-\,\overline{\Psi}_L M_\Psi\, \Psi_R
+ \text{h.c.},
\label{eq:lag}
\end{equation}
where $L$ and $H$ are the SM lepton and Higgs doublets.  The Lagrangian in Eq.~(\ref{eq:lag}) preserves both the baryon number $U(1)_B$ and the dark sector number $U(1)_\D$, the latter being spontaneously broken by the VEV of $\Phi$.\footnote{Eq.~(\ref{eq:lag}) induces a  $ \lambda_{H\Phi}|\Phi|^2 |H|^2$ coupling with $\lambda_{H\Phi} \sim y_{L,R}^2 y_\nu^2/(16 \pi^2)$. The values of $y_{L,R}$ and $y_\nu$ of our interest however imply $\lambda_{H\Phi} \ll 10^{-9}$, therefore not affecting our description of the PT, see the discussion in the end of Sec.~\ref{sec:SIpotential}.}
After electroweak and $U(1)_\D$ symmetry breaking, $H$ and $\Phi$ acquire VEVs $v_\EW$ and $v$ respectively, and the Yukawa couplings in Eq. (\ref{eq:lag}) generate the Dirac masses
\begin{equation}
M_\D = \frac{y_\nu \, v_\EW}{\sqrt2}, \qquad
M_{L,R} = \frac{y_{L,R}\, v}{\sqrt2}\,.
\end{equation}
The neutral fermion mass term in the basis
$\hat{\nu}=(\nu_L,\; N_R^{c},\; \Psi_L,\; \Psi_R^{c})$, is given by $\mathcal L=-\frac{1}{2}\overline{\hat{\nu}} \mathcal{M}\hat{\nu}^c +\hc$, with the mass matrix $\mathcal{M}$ defined as 
\begin{equation}
\mathcal{M}=
\begin{pmatrix}
0 & M_\D^T & 0 & 0\\[2pt]
M_\D & M_N & M_L^T & M_R^T\\[2pt]
0 & M_L & 0 & M_\Psi^T\\[2pt]
0 & M_R & M_\Psi & 0
\end{pmatrix}\,.
\end{equation}

\noindent Throughout this discussion we  work in the single generation case for the sake of notation and we assume real Yukawa couplings for simplicity. It is useful to block-decompose the mass matrix  $\mathcal{M}$ as
\begin{equation}
\mathcal M=
\begin{pmatrix}
0 & m_\D^T\\[3pt]
m_\D & M_H
\end{pmatrix}\,,
\end{equation}
with
\begin{equation}
M_H=
\begin{pmatrix}
M_N & M_L^T & M_R^T\\
M_L & 0 & M_\Psi^T\\
M_R & M_\Psi & 0
\end{pmatrix} ,\qquad
m_\D^T \equiv \bigl(M_\D^T\,,0\,,\,0\bigr)\,.
\end{equation}

We now assume a seesaw regime, in which the active–heavy mixing is perturbatively small,
\(\|\theta\| = \|M_H^{-1} m_\D\| \ll 1\).
This does not require a hierarchy among the individual entries of $M_H$, but rather that the eigenvalues of $M_H$ are parametrically larger than $m_\D$.
The mixing between the light SM neutrinos and the heavy fermions is given at leading order by
\begin{equation}
\theta = M_H^{-1} m_\D\,,
\label{eq:theta-definition}
\end{equation}
which upon inverting $M_H$ yields
\begin{equation} \label{eq:mixingangles}
\theta_{\nu N}=\frac{m_\D }{M_N -\frac{2 M_L M_R}{M_\Psi}},\hspace{0.3 cm}
\theta_{\nu\Psi_L}=-\frac{M_R \,m_\D}{M_N M_\Psi -2 M_L M_R }, \hspace{0.3 cm}
\theta_{\nu\Psi_R}=-\frac{M_L \,m_\D}{M_N M_\Psi-2 M_L M_R}\,
\end{equation}
at  first-order in \(m_\D\).
In the seesaw regime the light eigenvalue $\lambda$ of $\mathcal{M}$ is then  given by 
\begin{equation}
\lambda = -m_\D^{T} (M_H-\lambda I)^{-1} m_\D.
\label{eq:decomp}
\end{equation}
Since $\lambda = \mathcal O(m_\D^2)$, one may expand as
\begin{equation}
(M_H-\lambda I)^{-1}
= M_H^{-1} + \lambda M_H^{-2} + \mathcal O(\lambda^2).
\end{equation}
Inserting this into Eq. (\ref{eq:decomp}), 
solving for $\lambda$ to the first non-trivial order and defining the dimensionless ratio $\delta= 2 M_L M_R /(M_N M_\Psi)$, we get for the light neutrino masses
\begin{equation}
m_\nu = |\lambda|
=
\,\frac{m_\D^{2}}{M_N}\,\frac{1}{1-\delta}-\frac{m_\D^{4}}{M_N^{3}}\,
\frac{M_\Psi^{2}+M_L^{2}+M_R^{2}}{M_\Psi^{2}}\,
\frac{1}{(1-\delta)^{3}}
\;+\;
\mathcal{O}(m_\D^{6})
\label{eq:mnu-final}
\end{equation}

Interestingly, we note that the values of neutrino masses in the allowed range, $m_\nu \sim 0.1$~eV can be obtained for $M_N\sim 1-100$~MeV for sufficiently small values of $m_\D$, providing a possible explanation for the origin of neutrino masses.
The mixing between light neutrinos and $N, \Psi_{L,R}$ in Eq.~(\ref{eq:mixingangles}) is small, being controlled by $m_\D$. We can reexpress it in terms of light neutrino masses as $\theta_{\nu N}^2 \sim m_\nu / (M_N (1-\delta))$, $\theta_{\nu \Psi_{L,R}}= (M_{L,R}/M_\Psi) \theta_{\nu N}$.

From the $\Phi$-Yukawa terms in the Lagrangian Eq.~(\ref{eq:lag}), the dark Higgs $\phi$ can decay into two light neutrinos via the mixings $\theta_{\nu N}$ and $\theta_{\nu \Psi_{L,R}}$. This decay width is given by
\begin{equation} \label{eq:decayratemixing}
\Gamma (\phi \to \nu \overline{\nu}) = \frac{m_\phi}{8\pi} \,\theta_{\nu N}^2 |y_R \theta_{\nu\Psi_R}+y_L\theta_{\nu\Psi_L}|^2\,.
\end{equation}
We can obtain an order of magnitude estimate of this width by approximating $m_\D^2 \simeq m_\nu M_N (1-\delta)$ from Eq. (\ref{eq:mnu-final}) and plugging the relevant mixing angles from Eq. (\ref{eq:mixingangles}), with $y_L \approx y_R$. Conservatively neglecting possible enhancements from degeneracies and from including more generations, we then obtain
\begin{equation} \label{eq:decayratephi}
\begin{aligned}
\Gamma (\phi \to \nu \overline{\nu})
&\simeq \frac{m_\phi}{4\pi}\, y_{L,R}^4\,
\frac{m_\nu^2}{M_N^2}\,
\frac{v^2}{M_\Psi^2}\,
\mathcal{F}_\delta \simeq  6\times 10^{-22} \,\mathrm{GeV}\,
\left(\frac{y_{L,R}}{0.1}\right)^4
\left(\frac{m_\nu}{0.1\,\mathrm{eV}}\right)^2
\left(\frac{v}{100\,\mathrm{MeV}}\right)^3
\left(\frac{g}{0.6}\right)^2
\\ &
\left(\frac{10\,\mathrm{MeV}}{M_N}\right)^2
\left(\frac{10\,\mathrm{MeV}}{M_\Psi}\right)^2
\mathcal{F}_\delta\,,
\end{aligned}
\end{equation}
where $\mathcal F_\delta =\frac{1}{|1-\delta|^2}$ can be typically  of $\mathcal O(1)$. In the numerical estimate we normalize the parameters to representative benchmark values, in particular $y_{L,R}=0.1$, corresponding to values below the stability bound $y_{L,R}/g \lesssim 0.3$ required to prevent fermionic loop corrections from destabilizing the CW potential of Eq. (\ref{eq:oneloop}) \cite{Pascoli:2026tuu}. We thus see that, conservatively taking as reference timescale $\tau \ll 1~\mathrm{s} \sim  (6.5 \times 10^{-25} \ \mathrm{GeV})^{-1}$, decays before BBN can be achieved for our benchmark values.

Supernovae imply further constraints on the coupling of MeV scalars  to neutrinos~\cite{Heurtier:2016otg}, $g_\nu \phi \bar{\nu}\nu$, at the level of $g_\nu < 10^{-9}$ for the values of $m_\phi$ of our interest. Those constraints are satisfied in our model, where $g_\nu \approx y_{L,R}^2 (m_\nu/M_N) (v/M_\Psi) \lesssim 10^{-9}$ for our benchmark choices. Deleptonization arguments would lead to slightly stronger SN limits, but they depend on the details of the SN modeling and e.g. more recent ones lead to less stringent limits~\cite{Fiorillo:2022cdq}, so we do not consider them here.

We also need the heavy neutral lepton (HNL) mass eigenstates to decay before BBN, in order not to spoil its successful predictions. They result from the diagonalisation of $M_H$ and we denote them as $N_i$, $i=1,2,3$,  with masses $M_1<M_2<M_3$.
Under the assumption that $M_\Psi, M_N \geq m_\phi$, the relevant processes are $N_{j} \to \phi \, N_i$, $j>i$ if kinematically allowed, and $ N_{j} \to \phi \,\nu$. The former decays are typically very fast due to the large Yukawa coupling $y_{L,R}$. If they are not kinematically viable, and in any case for $N_1$, the decay into scalar and light neutrinos dominates. Focusing on $N_1$, if it is predominantly aligned along the $\Psi$ direction, $N_1 \simeq N_1^{\Psi}$, then this decay width can be estimated as
\begin{equation} \label{eq:decayratePsi}   
\Gamma(N_1^{\Psi} \to \phi \, \nu) \sim \frac{M_1 y_{L,R}^2}{8\pi}\, \theta_{\nu N}^2\simeq  1.3 \times 10^{-14} \,\text{GeV}\, \left(\frac{y_{L,R}}{0.1}\right)^2 \left(\frac{m_\nu}{0.1\,\mathrm{eV}}\right) \left(\frac{M_1}{10 \,\text{MeV}}\right)  \left(\frac{15 \,\text{MeV}}{M_N}\right)\sqrt{\mathcal F_\delta}\,,
\end{equation}
neglecting phase space effects possibly due to the masses of $N_i$ and $\phi$ being close.
On the other hand, if $N_1$ is mostly aligned along the $N$ direction, $N_1 \simeq N_1^{N}$, the decay width is suppressed by a $(\theta_{\nu \Psi }/\theta_{\nu N})^2$ factor with respect to Eq. (\ref{eq:decayratePsi}).
Therefore, for the benchmark values used above, the scalar $\phi$ and the heavy states $N_{1,2,3}$ typically decay well before BBN. 

From Eqs.~(\ref{eq:decayratemixing}) and~(\ref{eq:decayratePsi}), we notice that the values of the mixing angles $\theta_{\nu N}$ required for sufficiently fast scalar and heavy neutral lepton decays are compatible with neutrino masses. If, instead of a see-saw type I mechanism, we invoke an inverse or extended see-saw, adding multiple types of $N_R$, the mixing angles required by neutrino masses would be larger making the scalars and HNLs decay even faster. The required values of the mixing angles might be in principle tested at future HNL peak searches if the latter will very significantly improve their sensitivity and will be able to reach the see-saw mechanism parameter space, for reviews see e.g. \cite{Antel:2023hkf,deBlas:2025gyz}. We note that HNL decay bounds need to be reconsidered as, differently from the minimal case in which HNL mix only with active neutrino but do not have additional interactions, the heavy neutral leptons might predominantly decay invisibly, see Eq.~(\ref{eq:decayratePsi}), and, therefore, do not leave a visible signature in beam dump experiments. 

We can now compute the reheating temperature $\Treh$ upon the $U(1)_\D$ PT.
Its value is controlled by the interplay between the total decay width of the scalar $\Gamma_\phi$ and the Hubble parameter at $\Tnuc$. If $\Gamma_\phi>H (\Tnuc)$, then the scalars $\phi$ decay efficiently, otherwise they undergo a period of matter domination and $\Treh$ is suppressed. Assuming instantaneous reheating in the first case and reheating when $\Gamma_\phi \approx H$ in the second, one gets \cite{Ellis:2019oqb}
\begin{equation} \label{eq:reheating}
    \Treh \approx (1+\alpha_\GW)^{1/4} \Tnuc \,\text{min}\,\left(1,  \frac{\Gamma_\phi}{H (\Tnuc)}\right)^{1/2}\,,
\end{equation}
where the Hubble parameter at nucleation is given in the fast reheating scenario, with $g_*\sim 10$, by
\begin{equation} \label{eq:Hubble}
    H(\Tnuc) \approx H(\Treh) \simeq  2 \times 10^{-22} \text{GeV} \left (\frac {\Treh}{\text{20 MeV}}\right)^2 \,.  
\end{equation}
Comparing Eq.~(\ref{eq:decayratephi}) and Eq.~(\ref{eq:Hubble}) we find that  the condition $\Gamma_\phi > H(\Tnuc)$ is satisfied, for the typical reheating temperatures in our scenario (shown in Fig. \ref{fig:TResonance}) and for the benchmark values of the dark sector parameters discussed above. This implies that the fast reheating regime is viable, and therefore Eq.~(\ref{eq:reheating}) simplifies to $\Treh \simeq (1+\alpha_\GW)^{1/4} \Tnuc$, which we adopted in all our numerical computations.

\subsection{Thermal equilibrium between dark sector and SM }

We now investigate whether the SM and the dark sector are in thermal equilibrium in the temperature range relevant for the generation of the baryon asymmetry. We first need to address whether the sterile fermions $N_{\psi, N}$ thermalize upon the $\Phi$ PT. The relevant process are $\phi + \phi \to N_{\psi, N} + N_{\psi, N}$, induced by the Yukawa interactions in Eq. (\ref{eq:lag}). Assuming $M_{N_{\psi, N}} < T$ in our estimate, we then compute
\begin{equation}
\begin{aligned}
    \Gamma\left(\phi + \phi \to N_{\psi, N} + N_{\psi, N} \right) & \simeq n_\phi \,\sigma \left(\phi + \phi \to N_{\psi, N} + N_{\psi, N} \right) \simeq \frac{y_{L,R}^4}{4\pi}T \,
    \\ &
    \simeq 1.5 \times 10^{-7}\,\text{GeV}\,
\left(\frac{T}{20\,\text{MeV}}\right)
\left(\frac{y_{L,R}}{0.1}\right)^4 \,.
\end{aligned}
\end{equation}
This rate is orders of magnitude larger than the Hubble parameter, even if one took into account a possible suppression due to $M_{N_{\psi, N}}/T$ (which is at most of $\mathcal{O}(1)$ for our benchmark values), therefore we conclude that the sterile fermions are produced in thermal equilibrium after the PT.

We then proceed to estimate the rate of the dominant scattering processes connecting the SM and the dark sector.
We consider the $\nu + N_{\Psi} \xleftrightarrow{} N_{\Psi} + N_{\Psi} $ processes, where $N_{\Psi}$ denote the $N_i$ eigenstates aligned in the $\Psi$ direction. In the regime $M_\Psi < T < m_{V}$, integrating out the dark photon mediator yields the parametric estimate
\begin{equation}
    \sigma \left(\nu + N_{\Psi} \xleftrightarrow{} N_{\Psi}+ N_{\Psi}\right) \simeq\frac{g^4\,T^2}{\pi m_V^4} \theta_{\nu\Psi_{L,R}}^2\,\sim \frac{m_\nu} {2 \pi v^2 M_N} \,y_{L,R}^2 \frac{T^2}{M_\Psi^2}\sqrt{\mathcal F_\delta}\,.
\end{equation}
The scattering rate is then given by
\begin{align}
\Gamma\!\left(\nu + N_{\Psi} \xleftrightarrow{} N_{\Psi}+ N_{\Psi}\right)
&
\sim \frac{m_\nu} {2 \pi v^2 M_N} \,y_{L,R}^2 \frac{T^5}{M_\Psi^2}\sqrt{\mathcal F_\delta}
\simeq 
1.5 \times 10^{-16}\,\text{GeV}\,\left(\frac{m_\nu}{0.1\,\mathrm{eV}}\right)
\left(\frac{T}{20\,\text{MeV}}\right)^5
\left(\frac{y_{L,R}}{0.1}\right)^2 \nonumber \\
&\quad \times
\left(\frac{10\,\text{MeV}}{M_N}\right)  
\left(\frac{10\,\text{MeV}}{M_\Psi}\right)^2
\left(\frac{100\,\text{MeV}}{v}\right)^2 \sqrt{\mathcal F_\delta}\,\,.
\end{align}
Other processes, such as $\nu + N_{\Psi} \xleftrightarrow{} N_{\Psi} + N_{N}$, which is mediated by the dark Higgs $\phi$, yield a similar rate. By comparing the above rate with the Hubble parameter (\ref{eq:Hubble}), we then find that the SM and dark sectors are in thermal equilibrium during baryon asymmetry production.

\subsection{Washout processes and generation of the dark asymmetry}
\label{sec:washout}

The setup for the generation of the BAU that we described in Sec. \ref{sec:baryonasymmetry} assumes that a pre-existing dark baryon asymmetry in $\chi$ is processed into a visible baryon asymmetry by $\chi \to n$ oscillations after the $\Phi$ PT. However, the relevant operator $\mathcal{O}_{\rm eff}$ for the production of the $\chi \to n$ oscillations, given by Eq.~(\ref{eq:effectiveop7}), could be kept in thermal equilibrium above the QCD scale by processes such as $\phi \chi \to udd$, as noted in \cite{Bringmann:2018sbs}. In this regime, chemical equilibrium tends to reprocess part of the $\chi$ asymmetry into a visible baryon asymmetry, potentially spoiling our mechanism. We then need to estimate the decoupling temperature  $\Tfo$ of $\mathcal{O}_{\rm eff}$. For $T \gg m_\chi$ we can approximate all the particles involved as massless, hence the scattering rate $\Gamma$ can be estimated as 
\begin{equation}
\Gamma (\phi \chi \to udd)\simeq n_\chi \sigma_{\phi \chi \to udd }  \simeq  \frac{1}{(4\pi)^3}\frac{T^7}{\Lambda_7^6}\,.
\end{equation}
Therefore, by inverting Eq. (\ref{eq:deltam}) and by requiring $\Gamma < H$ at decoupling, we find $\Tfo \sim  \Lambda_7 \left(\Lambda_7/M_\Pl \right)^{1/5} \sim 11 \text { GeV}\, \left(v/ 100 \text{ MeV}\right)^{2/5} (10^{-10}{\rm\,MeV}/\delta m)^{2/5}.$
The $\mathcal{O}_{\rm eff}$-induced washout can then be avoided if the Universe reheats after inflation at $T_\text{infl} \leq \Tfo$.

Another possibility to avoid that $\mathcal{O}_{\rm eff}$ washes out the DM and baryon asymmetries, even for reheating temperatures larger than $\Tfo$ for which it would be in equilibrium, is that also the scalar particles are asymmetric and have a chemical potential $\mu_\phi$ satisfying $\mu_\chi = \mu_\phi$.
The latter condition can be enforced dynamically, for instance by introducing a heavy massive Majorana fermion $\psi'$~\cite{Bringmann:2018sbs}, decaying via $\psi' \to \chi \phi$, which by construction generates the same dark asymmetry in $\chi $ and $\phi$.\footnote{The interactions inducing $\psi' \to \chi \phi$ may also induces a Majorana mass term for $\chi$ after $\phi$ acquires a VEV, in which case they would lead to $\chi$-$\bar{\chi}$ oscillations. However, one can set the relevant parameters such that the induced mass splitting is extremely small and such oscillations are efficiently damped by scattering processes down to temperatures well below those relevant for baryogenesis~\cite{Bringmann:2018sbs}.}
We point out that the decays $\psi' \to \chi \phi$  could actually be the ones responsible for the generation of the initial dark asymmetry in $\chi$, whose origin we had not specified so far. Studying the generation of the $\chi$ and $\phi$ asymmetries from these decays does not appear to pose particular challenges, as it is a standard asymmetry-generation mechanism and it happens in a dark sector. Therefore we do not expand on it further in this paper, but we just stress that achieving $\mu_\chi = \mu_\phi$  would not constitute an additional ingredient in our model.

We now discuss how $\mu_\chi = \mu_\phi$ evolves as the universe cools down. As we assume no further number changing interactions, $\mu_\chi=\mu_\phi$ remains effectively frozen-in after its creation and no baryon asymmetry is induced before the onset of $\chi \to n$ oscillations. After the PT only a real scalar degree of freedom survives in the scalar $\phi$, that cannot store any asymmetry any longer. The asymmetry in $\phi$ is then transferred to the other light particles charged under $U(1)_\D$, i.e. the HNLs. They then transfer it to neutrinos via decays like Eq.~(\ref{eq:decayratePsi}), but as much large asymmetries in the neutrino sectors are still allowed by data~\cite{Domcke:2025jiy} this does not constitute a problem.

The same operator $\mathcal{O}_{\rm eff}$ of Eq.~(\ref{eq:effectiveop7}) could in principle induce efficient washout processes of the baryon and DM asymmetries after QCD confinement. We decompose $\Phi=(v+\phi+i\phi_i)/\sqrt{2}$, which upon $U(1)_\D$ breaking induces
$\mathcal L\supset -\delta m\,\bar n\chi - y_{n\chi}\,\phi \,\bar n\chi+\hc$ with
$y_{n\chi}\simeq \delta m/v \sim 10^{-12}\,(\delta m/10^{-10}{\rm\,MeV})(100{\rm\,MeV}/v)$. Since $m_\phi \gg |\Delta m|$, decays $n \to \chi \phi$ are not kinematically allowed, and the asymmetry is not washed out.
We therefore conclude that scalar-induced washout effects are negligible and do not spoil our mechanism to generate the baryon and DM abundance.

\section{Conclusions and outlook}
\label{sec:summary}

The recent observation of a stochastic gravitational wave background (SGWB) at nanohertz frequencies by the pulsar timing arrays (PTAs) has opened a new window on early Universe physics. If interpreted as originating from a cosmological first-order phase transition (FOPT), the signal points to new physics at temperatures of order $\mathcal{O}(100)\,\mathrm{MeV}$. This naturally raises the question of what underlying particle physics dynamics could operate at such a scale.

In this work, we have proposed the possibility that this scale is not accidental, but linked to the baryon-dark matter (DM) coincidence problem, namely the puzzling similarity of their energy densities $\rho_{\rm DM} \simeq 5 \rho_b$. In particular, we have revisited a darkogenesis scenario proposed in~\cite{Bringmann:2018sbs}, in which the baryon asymmetry of the Universe (BAU) is generated via resonant neutron-DM oscillations and the mixing responsible for these oscillations is induced by the vacuum expectation value of a scalar field.
We have altered the model of~\cite{Bringmann:2018sbs} by making the PT strongly first-order, which allowed us to link the associated FOPT with the baryon-DM coincidence, and where the scale of $\mathcal{O}(100)\,\mathrm{MeV}$ is required  by successful baryogenesis.\footnote{
\label{foot:PTA_BAU}
The connection between PTA GW from a first-order PT, DM and the BAU has also been explored in~\cite{Feng:2025wvc}, where co-genesis takes place in the UV and a dark PT then generates GW, and in~\cite{Fujikura:2024jto}, where a dark asymmetry is transferred to the baryons via the portal $\chi u d d$ (so not via neutron-DM oscillations) and GWs arise from a PT that also generates the dark asymmetry via textures~\cite{Fujikura:2024jto}. While interesting, these models can operate also at larger scales and therefore do not address the question of ``Why 100 MeV?''.
Refs.~\cite{Girmohanta:2025wcq,Girmohanta:2026eqg} also link DM and the BAU via $\chi u d d$ and feature a PT, and~\cite{Girmohanta:2026eqg} further links the PT scale to (specific) choices for the UV completion of $\chi u d d$. However,~\cite{Girmohanta:2025wcq,Girmohanta:2026eqg} do not check if their PT fits the PTA signal, which as we have worked out can be an extremely constraining requirement, so their connection between DM, the BAU and observable nHz GW still lacks the burden of proof.
}
Requiring the same underlying dynamics to account for both baryogenesis and a PTA nHz GW signal significantly restricts the parameter space of the model. In particular, the scale of the PT is not treated as a free parameter to be fitted to the PTA signal, but is constrained by the requirement of successful baryogenesis. We identify a non-trivial region in parameter space where (i) successful generation of the BAU, (ii) compatibility with the observed nanohertz GW signal and (iii) consistency with cosmology and astrophysical constraints are simultaneously satisfied, see Fig.~\ref{fig:parameterspace}.

The proposed scenario has several phenomenological implications and is testable through multiple complementary avenues.
On the astrophysical side, the scenario is compatible with neutron star (NS) masses up to $\sim 2\,M_\odot$, within current uncertainties on the NS equation of state, and predicts DM self-interactions that could give observable effects in the distribution of matter on large scales.
Collider searches, in particular monojet plus missing energy signatures, are sensitive to the effective operator responsible for neutron-DM mixing. 
A consistent cosmological history motivates a minimal extension of the model, that includes heavy neutral leptons with masses around $\mathcal{O}(10\text{-}100)\,\mathrm{MeV}$ and fast decays to avoid BBN, which generate neutrino masses via see-saw and open additional experimental probes.
Future cosmological measurements, in particular CMB ones, are expected to fully test the parameter space where this baryogenesis model could also explain the origin of the PTA signal.

\medskip

Looking ahead, our conclusions rely on the modeling of GW production from strongly supercooled FOPTs in the non-runaway regime. Different templates for the GW signal can lead to $\mathcal{O}(1)$ variations in both the peak frequency and amplitude (see Fig.~\ref{fig:gwtemplates}). This introduces a theoretical uncertainty in extracting the model parameters from the observed GW, and highlights the need for improved understanding of GW production in this regime.
Such understanding would also strengthen the solidity of the interplay of the preferred PTA region with other tests of the model.

Concerning model-building aspects, in our opinion the main value of our proposal lies in the fact that it is (to our knowledge) the first one where the PT scale of $\mathcal{O}(100)\,\mathrm{MeV}$ is not plugged in by hand, but demanded by reasons independent of the PT. On the other hand, the baryogenesis mechanism that we employ relies on extremely close values for $m_n$ and $m_\chi$, whose origin is not specified here nor in~\cite{Bringmann:2018sbs}. Demanding the PT to be strongly first-order also requires the $\phi$-Higgs portal to be very small. While neither small numbers pose a problem of technical naturalness, the model would arguably be more appealing if their smallness was explained.
This adds to the motivation to explore further connections between the nHz GW signal and the generation of the baryon asymmetry at~$\mathcal{O}(100)\,\mathrm{MeV}$.
Directions worth exploring include the generation of GW in mesogenesis~\cite{Elor:2018twp,Baruch:2026iwn} and the extension to small VEVs of the mechanisms of BAU generation from ultrarelativistic walls, such as~\cite{Baldes:2021vyz,Azatov:2021irb} (which work only at much larger VEVs by construction, but also at small VEVs if the PT is confining~\cite{Dichtl:2023xqd}) and~\cite{Bhusal:2025lvm} (if the mechanism will survive the inclusion of CP violation and non-runaway effects). 

\medskip

Overall, our results demonstrate that combining GW observations with particle physics and cosmological considerations not only can sharpen the interpretation of the PTA signal, but also point to a motivated and testable target for new physics at sub-GeV scales.

\section*{Acknowledgements}

FS thanks Mikael Chala for useful discussions on calculating the PT parameters at higher orders, and Thomas Konstandin for useful discussions on gravitational waves from ultrarelativistic walls. We thank Matteo Zandi and Salvador Rosauro-Alcaraz for discussions on the gravitational wave templates.
The work of SP and FS have  been partly funded by the European Union under the Horizon Europe’s Project: 101201278 – DarkSHunt - ERC - 2024 ADG.
JN is supported in part by the Strategic Research Program High-Energy Physics of the Research Council of the Vrije Universiteit Brussel, and by the FWO-IRI I000725N of the Fonds Wetenschappelijk Onderzoek (FWO).

\appendix

\appendix 
\section{Running Couplings}
\label{app:RGE}

The relevant one-loop beta functions in the conformal dark $U(1)$ model are given by:\\ 
\begin{equation}
    \beta_g= \frac{g^3}{48 \pi^2}, \hspace{2 cm} \beta_\lambda= \frac{3g^4+10\lambda^2-6\lambda g^2}{8\pi^2}\,. \label{eq:betafunction}
\end{equation}
The former can be easily integrated, yielding 
\begin{equation}\label{eq:gaugecoupling}
    g^2(\mu) =\frac{g_0^2}{1-\frac{1}{24\pi^2}g_0^2 \log{\frac{\mu}{\mu_0}}},
\end{equation}
where we choose $\mu_0 =v$ and define $g_0=g(\mu_0)$. \\
The beta function of the quartic coupling can be integrated by defining $R=\lambda/g^2$, which by using Eq. (\ref{eq:gaugecoupling}) leads to the following expression: 
\begin{equation}
g^2 \frac{dR}{dg^2}= 9 -19 R+ 30R^2\,.
\end{equation}

This can be integrated analytically, yielding
\begin{equation}
    \lambda (\mu)=\frac{g^2(\mu)}{60}\left[19 +\sqrt{719}\left(\tan\left(\frac{\sqrt{719}}{2}\log(g^2(\mu))+\theta (g_0)\right)\right)\right]\,,
    \label{eq:lambda_mu}
\end{equation}
where $\theta(g_0)$ is chosen such that 
\begin{equation}
\left.\frac{d V_{\rm eff}}{d \phi}\right|_{\phi = v,\ \mu = v} = 0 \, .
\end{equation}

Finally, the scalar field renormalization is given by $\phi \to \sqrt{Z_\phi(\mu)} \phi_b$, where $\phi_b$ the unrenormalized field, and $Z_\phi(\mu)$ is given by
\begin{equation} \label{eq:waveren}
    Z_\phi(\mu)= \text{Exp}\left[-2 \int_0^t dt' \gamma_\phi\left(g(t')\right)\right]\,,
\end{equation}
being $\gamma_\phi=-3g^2/(16\pi^2)$ the one-loop anomalous dimension of the scalar field. Upon integration of Eq. (\ref{eq:waveren}) and plugging in Eq. (\ref{eq:gaugecoupling}) one obtains 
\begin{equation}
    Z_\phi(\mu)=\text{Exp}\left[-9 \left(\log \left(1-\frac{g_0^2}{24 \pi^2} \log \frac{\mu}{\mu_0}\right)\right)\right]\,.
\end{equation}

\section{GW spectrum templates}
\label{app:GWtemplates}

\paragraph{Envelope approximation}

The SGWB generated during a FOPT receives contributions from bubble collisions and from sound waves in the plasma. Here we report the relevant equations for the former within the envelope approximation \cite{Jinno:2016vai}, adopted by the NANOGrav collaboration in \cite{NANOGrav:2023hvm}. The GW spectrum energy density is given by 
\begin{align}
h^2\Omega_{b}(f)
&=
\mathcal{D}\,\tilde{\Omega}_{b} \left(\frac{\alpha_\GW}{1+\alpha_\GW}\right)^2
(H_\ast R_\ast)^2\,
S_b\!\left(\frac{f}{f_b}\right)\,.
\end{align}
Here $\mathcal{D} \simeq 1.67\times 10^{-5}\,(100/g_\ast)^{1/3}$ is the redshift factor, $\tilde{\Omega}_b = 0.0049$ \cite{Jinno:2016vai}, $R_\ast$ is the mean bubble separation, and $H_\ast R_\ast= (8\pi)^{1/3}v_w/\beta_H$. The peak frequency observed today is given by 
\begin{equation}
f_b
\simeq
48.5\,{\rm nHz}\,
g_\ast^{1/2}
\left(\frac{g_{\ast,s}^{\rm eq}}{g_{\ast,s}}\right)^{1/3}
\left(\frac{\Treh}{1\,{\rm GeV}}\right)
\frac{f^\ast_bR_\ast}{H_\ast R_\ast},
\end{equation}
where the entropic and relativistic degrees of freedom $g_{\ast,s}$ and $g_\ast$ are evaluated at $T=\Treh$, while $g_{\ast,s}^{\rm eq}$ is evaluated at matter-radiation equality. The quantity $f_b^\ast \simeq 0.58/R_\ast$ is the peak frequency at the time of GW production. Finally, the spectral shape $S_b$ is modeled by a broken power law
\begin{equation}
S_b(x) =\frac{1}{\mathcal N}
\left( b\,x^{-a/c} + a\,x^{\,b/c} \right)^{-c},
\end{equation}
where $a$ and $b$ determine the low and high-frequency slopes, $c$ controls the width of the peak and $\mathcal N$ ensures the proper normalization. From the NANOGrav Bayes posterior we set $a=b=c=2$ in our computation \cite{NANOGrav:2023hvm}.

\paragraph{Dissipative bulk flow} 

A recent analysis \cite{Lewicki:2025hxg} improves over the original bulk flow model \cite{Konstandin:2017sat} in the computation of the GW spectrum by taking into account dissipative effects of the shells upon the bubble collisions and the expansion of the Universe at GW production. The GW spectrum energy density can be parametrized as 

\begin{equation}
h^2 \Omega_\GW(f)=h^2 \bar{\Omega}_\GW^{(0)}\, S(f,f_\text{reh}),
\end{equation}
where the amplitude today, accounting for the redshift of the GW energy density, is given by
\begin{equation} \label{eq:redshiftGW}
h^2 \bar{\Omega}_{\mathrm{GW}}^{(0)} \simeq 1.6\times10^{-5}
\bigg(\frac{1}{\beta_H}\bigg)^2 \left(\frac{\alpha_\GW}{1+\alpha_\GW}\right)^2 \left(\frac{g_\ast}{100}\right)
\left(\frac{g_{\ast s}}{100}\right)^{-4/3} A (\beta_H)\,,
\end{equation}
with $g_\ast$ ($g_{\ast s}$) the effective number of energy (entropy) degrees of freedom at $T=\Treh$ and $A(\beta_H)$ the fitted amplitude at production. The dependence of the overall normalization on the inverse PT duration parameter $\beta_H$ can be approximated as
\begin{equation}
A(\beta_H)\simeq 0.06\left[1+0.8\left(1-e^{-\frac{1}{\sqrt{\beta_H}}}\right)\right]\,.
\end{equation}
The spectral shape is described by a broken power-law of the form
\begin{equation}
S(f,f_\text{reh})=
\frac{(a+b)^c}
{\left(b\, \left(\frac{f}{f_\text{reh}}\right)^{-\frac{a}{c}}+a\, \left(\frac{f}{f_\text{reh}}\right)^{\frac{b}{c}}\right)^{c}},
\end{equation}
where $f_\text{reh}$ is the peak frequency at production and $a\simeq b\simeq c \simeq 2$. The observed frequency today is related to that at production by
\begin{equation}
f_0 \simeq 2.6\times10^{-8}\,\mathrm{Hz}
\left(\frac{\Treh}{\mathrm{GeV}}\right)
\left(\frac{g_\ast}{100}\right)^{1/2}
\left(\frac{g_{\ast s}}{100}\right)^{-1/3}
\frac{2\pi f}{a_\text{reh} H_\text{reh}}\,.
\end{equation}
The effect of cosmic expansion on supercooled PTs leads to a shift of the peak frequency, which can be approximated as
\begin{equation}
\frac{2\pi f_\text{reh}(\beta_H)}{a_\text{reh} H_\text{reh}}
\simeq
0.7\,\beta_H
\left[1+1.8\beta_H^{-1.2}\right]\,.
\end{equation}
Finally, at frequencies below the inverse horizon scale, a causality-driven tail emerges, characterized by
\begin{equation}
\frac{2\pi f_H(\beta_H)}{a_\text{reh} H_\text{reh}}
\simeq
1-0.6\,\beta_H^{-0.67},
\end{equation}
corresponding to a low-frequency behavior $\Omega_\GW(f)\propto f^3$ for $f<f_H$.

\paragraph{GWs from sound waves and turbulence}

Assuming relativistic bubble walls with $v_w\simeq1$, and terminal velocity regime, as is the case in our setup, the efficiency coefficient for GW  production from sound waves in the plasma can be approximated as \cite{Espinosa:2010hh}
\begin{equation}
\epsilon_{\rm sw} =
\frac{\alpha_\GW}
{0.73 + 0.083\sqrt{\alpha_\GW} + \alpha_\GW}\,.
\end{equation}
The GW energy density from sound waves at the time of production is then given by \cite{Ellis:2019oqb}
\begin{equation}
\Omega_{\rm sw,\ast}(f_\ast)
=
0.38\,(H_\ast R_\ast)\,(H_\ast \tau_{\rm sw})
\left(\frac{\epsilon_{\rm sw}\alpha_\GW}{1+\alpha_\GW}\right)^2
\left(\frac{f_\ast}{f_{\rm sw}}\right)^3
\left[1+\frac{3}{4}\left(\frac{f_\ast}{f_{\rm sw}}\right)^2\right]^{-7/2},
\end{equation}
where $\tau_{\rm sw}$ denotes the duration of the sound wave period. The spectrum peaks at the frequency $f_{\rm sw} = 3.4/((v_w-c_s)R_\ast)$, with the speed of sound in the plasma $c_s=1/\sqrt{3}$. The factor $H_\ast\tau_{\rm sw}$ accounts for the finite lifetime of the acoustic source, which is taken to be $\tau_{\rm sw} \equiv \min\!\left[\frac{1}{H_\ast},\,\frac{R_\ast}{\bar{U}_f}\right]$,  being $\bar{U}_f$ the root-mean-square fluid velocity, which can be approximated by \cite{Hindmarsh:2015qta} $\bar{U}_f^2 \simeq
\frac{3}{4}\frac{\alpha_\GW}{1+\alpha_\GW}\epsilon_{\rm sw}$.
If the sound wave period is shorter than a Hubble time, part of the bulk kinetic energy cascades into magnetohydrodynamic turbulence, which also sources a GW signal. The corresponding contribution to the GW spectrum at production is given by \cite{Caprini:2009yp}
\begin{equation}
\Omega_{\rm turb,\ast}(f_\ast)
=
6.8\,(H_\ast R_\ast)\,(1-H_\ast\tau_{\rm sw})
\left(\frac{\epsilon_{\rm sw}\alpha_\GW}{1+\alpha_\GW}\right)^{3/2}
\frac{\left(f_\ast/f_{\rm turb}\right)^3}
{\left[1+8\pi f_\ast/H_\ast\right]
\left[1+\left(f_\ast/f_{\rm turb}\right)\right]^{11/3}},
\end{equation}
which peaks at the frequency $f_{\rm turb} = 3.9/((v_w-c_s)R_\ast)$.
The total GW spectrum at production is given as $\Omega_{\rm GW,\ast}=\Omega_{\rm sw,\ast}+\Omega_{\rm turb,\ast}$ and the present-day GW is obtained by redshifting the frequencies and amplitudes according to the cosmological expansion.

\begin{figure}[H]
        \centering
        \includegraphics[width = 0.8\textwidth]{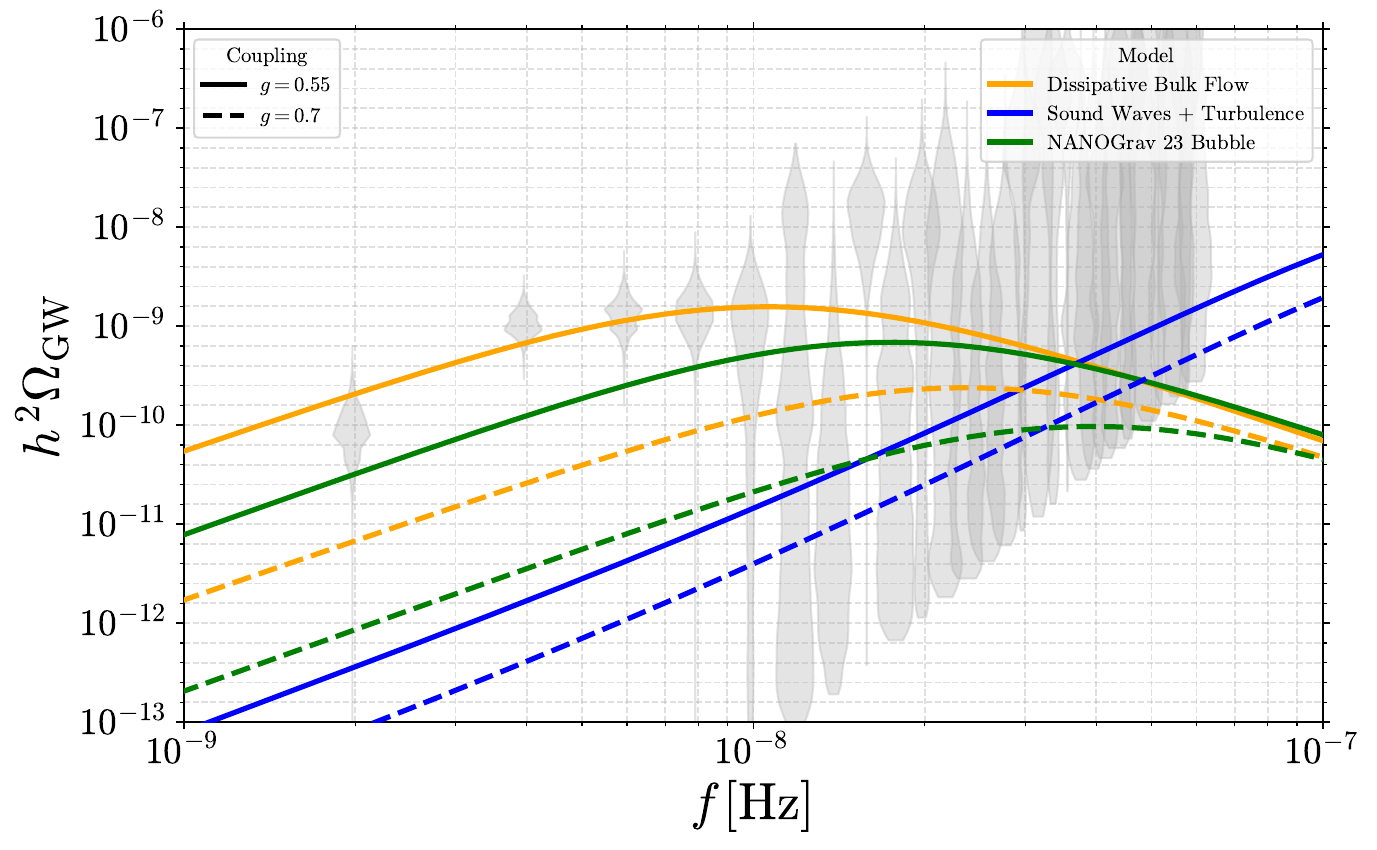}
        \caption{ Predicted gravitational wave spectrum, using the dissipative bulk flow model \cite{Lewicki:2025hxg} (yellow), the envelope approximation \cite{Jinno:2016vai} (green) and sound waves plus turbulences \cite{Ellis:2019oqb} (blue) for the relevant PT parameters $\alpha_\GW$, $\beta_H$, $\Tnuc$, $\Treh$ computed for benchmark values $g=0.55$ (solid), $g=0.7$ (dashed) and setting the scalar field VEV at $v=100$ MeV. The gray regions correspond to the NANOGrav's violin plot \cite{NANOGrav:2023gor}.}
                \label{fig:gwtemplates}

\end{figure}

We show in Fig. \ref{fig:gwtemplates} a comparison of the predicted GW signal for the templates discussed above for the benchmark parameters $g=0.55,0.7$ and $v=100$ MeV. We observe that the choice of template has a significant impact, leading to $\mathcal{O}(1)$ differences in both the predicted peak frequency and the overall normalization of the GW signal.

\bibliography{Baryogenesis_bib}
\bibliographystyle{JHEP}

\end{document}